\documentclass[11pt]{article} 

\usepackage{graphicx}
\usepackage{subfigure}
\usepackage{color}
\usepackage{epsfig}
\usepackage{amssymb,url}
\usepackage{latexsym}
\usepackage{relsize}

\newcommand {\C} {{\rm I\kern-5.5pt C}}

\newcommand{\bP}[1]{{\mathbb{P}}\left[{#1}\right]}
\newcommand{\bE}[1]{{\mathbb{E}}\left[{#1}\right]}

\newcommand{\1}[1]{{\bf 1}\left[#1\right]}       

\newcommand{\fsquare}{\vrule height6pt width7pt depth1pt}   
\newcommand{\myproof}{{\hfill \\ \bf Proof. \ }}           
\newcommand{\myendpf}{\hfill\fsquare \\[0.1in]}             

\newtheorem{theorem}{Theorem}[section]
\newtheorem{definition}[theorem]{Definition}
\newtheorem{lemma}[theorem]{Lemma}
\newtheorem{proposition}[theorem]{Proposition}
\newtheorem{corollary}[theorem]{Corollary}
\newtheorem{conjecture}[theorem]{Conjecture}

\begin{document}
\title{Towards $k$-connectivity of the random graph induced by a pairwise key predistribution scheme with unreliable links
\thanks{A short version of this paper (without any proofs) will be presented at
IEEE International Symposium on Information Theory, (ISIT 2014), Honolulu (HI).}}

\author{
Faruk Yavuz \and Jun Zhao \and Osman Ya\u{g}an \and Virgil Gligor\\
{\tt \{fyavuz,junzhao,oyagan,gligor\}@andrew.cmu.edu} \\
Department of Electrical and Computer Engineering and CyLab\\
Carnegie Mellon University, Pittsburgh, PA 15213.\\
}
\maketitle
\begin{abstract}
\normalsize We study the secure and reliable connectivity of wireless sensor networks. 
Security is assumed to be ensured by the random pairwise key predistribution 
scheme of Chan, Perrig, and Song, and unreliable wireless links are represented by independent on/off channels. 
Modeling the network by an intersection of a random $K$-out graph and an Erd\H{o}s-R\'enyi graph, we present scaling conditions (on the number of nodes, the scheme parameter $K$, and the probability of a wireless channel being on) such that the resulting graph contains no nodes with degree less than $k$ with high probability, when the number of nodes 
gets large. Results are given in the form of zero-one laws and are shown to improve the previous results by Ya\u{g}an and Makowski on the absence of isolated nodes (i.e., absence of nodes with degree zero). Via simulations, the established zero-one laws are shown to hold also for the property of $k$-connectivity; i.e., the property that graph remains connected despite the deletion of any $k-1$ nodes or edges.
\end{abstract}

\section{Introduction}
\label{sec:intro}
\label{sec:Introduction}

\subsection{Motivation and Background}
Wireless sensor networks (WSNs) are
 distributed collection of small sensor nodes that gather security-sensitive data and 
 control security-critical operations in a wide range of industrial, home and business applications \cite{Akyildiz}. 
 Many applications require deploying sensor nodes in hostile environments 
 where an adversary can eavesdrop sensor communications, 
 and can even capture a number of sensors and surreptitiously 
 use them to compromise the network. Therefore, cryptographic protection 
 is required to secure the sensor communication as 
 well as to detect sensor capture and to revoke the compromised keys. 
 Given the limited communication and computational 
 resources available at each sensor, security is expected to be a key challenge in WSNs
 \cite{virgil,ChanPerrigSong,WangAttebury}.

Random key predistribution is one of the approaches proposed
in the literature for addressing security challenges in
resource constrained WSNs.
The idea of randomly
assigning secure keys to the sensor nodes prior to network
deployment was first introduced by Eschenauer and Gligor
\cite{virgil}. Following their original work, 
a large number of key predistribution schemes
have been proposed; see the survey articles
\cite{WangAttebury,XiaoSurvey} (and references therein).

Here we consider the random pairwise key predistribution
scheme proposed by Chan et al. in \cite{ChanPerrigSong}:
Before deployment, each of
the $n$ sensor nodes is paired (offline) with $K$ distinct nodes
which are randomly selected from amongst all other nodes. For 
each sensor and any sensor paired to it,
a unique (pairwise) key is generated and stored in their memory modules 
along with their ids. Two nodes can then secure an existing wireless
communication link if at least one of them is paired
to the other so that the two nodes have at least one
pairwise key in common. Precise implementation details are given
in Section \ref{sec:Model}. 

Let $\mathbb{H}(n;K)$ denote the undirected random graph on the vertex set
$\{ 1, \ldots , n \}$ where distinct nodes $i$ and $j$ are
adjacent if they have a pairwise key in common as described earlier;
this random graph models the random pairwise predistribution scheme
under {\em full visibility} (whereby all nodes have a wireless link in between).
The random graph 
$\mathbb{H}(n;K)$ is known in the literature
on random graphs as the random $K$-out graph
\cite{Bollobas,FennerFrieze,JansonLuczakRucinski}, and is typically defined in the following
equivalent manner: For each of the $n$ vertices assign exactly $K$ arcs to
$K$ distinct vertices that are selected
uniformly at random, and then ignore the orientation of the arcs. 
Several properties of this graph have been recently analyzed
by Ya\u{g}an and Makowski \cite{YM_WiOpt2011, YM_ISIT2012,
YM_PerfEval, YaganMakowskiPairwise}. 

Recently, there has been a significant interest \cite{Krishnan,YaganMakowskiPER,Yagan_onoff,ZYG_ISIT2013,ZhaoYaganGligor} 
to drop the full visibility assumption and to
model and analyze 
random key predistribution 
schemes under more realistic situations that account for 
the possibility that communication links between nodes
may not be available -- This could occur due to the presence of
physical barriers between nodes or because of harsh environmental
conditions severely impairing transmission. With this in mind, several 
authors \cite{YaganMakowskiPER,Yagan_onoff,ZYG_ISIT2013,ZhaoYaganGligor} 
have started with a simple communication model where
wireless links are represented by independent channels that
are either on (with probability $p$) or off (with probability $1-p$). This suggests
an overall modeling framework that is constructed by {\em
intersecting} the random $K$-out graph $\mathbb{H}(n;K)$, with an
Erd\H{o}s-R\'enyi (ER) graph $\mathbb{G}(n;p)$ \cite{Bollobas}. 

\subsection{Contributions}
In this paper, we initiate an analysis towards
the $k$-connectivity for
the resulting intersection graph $\mathbb{H \cap G}(n;K,p)$. A
network (or graph) is said to be $k$-connected 
if its connectivity is preserved despite the failure 
of any $(k-1)$ nodes or links \cite{PenroseBook}. 
Therefore,
the property of $k$-connectivity provides a guarantee of network reliability 
against the possible failures of sensors or links due
to adversarial attacks or battery depletion; a much needed property given the 
key application areas of sensor networks such as health monitoring, battlefield surveillance,
and environmental monitoring.
Finally, $k$-connectivity has important benefits in {\em mobile} wireless sensor networks. 
For instance, if a network is known to be $k$-connected, then any $k-1$ nodes in the network 
are free to move anywhere in the network while the rest of the network remains at least 1-connected. 

Our main result is a zero-one law for the property that the minimum node degree of  $\mathbb{H \cap G}(n;K,p)$
is at least $k$. Namely, we present scaling conditions on the parameters $p$ and $K$ with respect to $n$, such that the resulting graph contains no nodes with degree less than $k$ with probability approaching to zero, or one, respectively, as the number of nodes $n$ gets large. The established results already imply the zero-law for the $k$-connectivity, since a graph can not be $k$-connected unless all nodes have degree at least $k$. Further, in most random graph models
in the literature, including ER graphs, random geometric graphs \cite{PenroseBook}, 
and random key graphs \cite{ZhaoYaganGligor}, the conditions that ensure $k$-connectivity coincide with those ensuring minimum node degree
to be at least $k$. This is often established by showing the improbability of a 
graph being {\em not} $k$-connected
when all nodes have at least $k$ neighbors.
Here, we demonstrate this phenomenon via simulations which suggest that  
our zero-one laws hold also for the property of $k$-connectivity. 

Furthermore, our results with $k=1$
constitute an improvement of the previous results by 
Ya\u{g}an and Makowski \cite{YM_ICC2011,YaganMakowskiPER} 
on the absence of isolated nodes 
(i.e., absence of nodes with degree zero) in $\mathbb{H \cap G}(n;K,p)$. Namely, we show that the threshold for 
absence of isolated nodes (which is also the threshold for 1-connectivity) characterized in \cite{YM_ICC2011,YaganMakowskiPER} 
is not valid unless the limit $\lim_{n \to \infty} p_n \in [0,1]$ exists, a condition that was enforced throughout in \cite{YM_ICC2011,YaganMakowskiPER}.
Instead, our main result indicates a new threshold function which does not require the existence of $\lim_{n \to \infty} p_n$. More importantly,
 we show that the new threshold function is {\em stronger} in that it indicates a sharper transition of the graph
$\mathbb{H \cap G}(n;K,p)$ (as the parameters $K$ and $p$ increase) from having at least one isolated to having no isolated nodes almost surely; see Section \ref{subsec:CompYM} for details. We believe that the precise characterization of the threshold for absence of isolated will also pave the way
to improving the results of \cite{YM_ICC2011,YaganMakowskiPER} for 1-connectivity of $\mathbb{H \cap G}(n;K,p)$.

Finally, our main contributions include a key confinement result that not only eases the proof of our main result, but is likely to play a key 
role in studying any {\em monotone increasing}\footnote{A graph property
is called monotone increasing if it holds under the addition of
edges in a graph. \label{fnote:gr_prop}} property of the graph $\mathbb{H \cap G}(n;K_n,p_n)$; e.g., $k$-connectivity, existence of certain subgraphs, etc.
In a nutshell, this confinement
result shows that when seeking results for the asymptotic $k$-connectivity of $\mathbb{H \cap G}(n;K_n,p_n)$ with the parameters $K$ and $p$ scaled with number of nodes $n$, we can restrict our attention to a subclass of structured scalings (referred throughout as admissible scalings). In other words, we 
show that the aforementioned results (and others in the same vein) need only be established for such strongly admissible scalings. See Section \ref{subsec:confining} for details of the confinement argument, followed in Section \ref{subsec:useful} by its several useful consequences that arise in our context.

\subsection{Notation and conventions}
A word on the notation: All statements involving limits
are understood with $n$ going to infinity. The random
variables (rvs) under consideration are all defined on the same
probability triple $(\Omega, {\cal F}, \mathbb{P})$. Probabilistic
statements are made with respect to this probability measure
$\mathbb{P}$, and we denote the corresponding expectation operator
by $\mathbb{E}$. The indicator function of an event $E$ is
denoted by $\1{E}$. Distributional equality is denoted by 
$=_{st}$. In comparing
the asymptotic behaviors of the sequences $\{a_n\},\{b_n\}$, 
we use
$a_n = o(b_n)$,  $a_n = O(b_n)$, $a_n = \Omega(b_n)$, and
$a_n = \Theta(b_n)$, with their meaning in 
the standard Landau notation. Namely, we write 
$a_n = o(b_n)$ as a shorthand for the relation
 $\lim_{n \to \infty} \frac{a_n}{b_n}=0$, whereas $a_n = O(b_n)$
means that there exists $c>0$ such that $a_n \leq c b_n$ for all
$n$ sufficiently large. Also, we have $a_n = \Omega(b_n)$ if
$b_n=O(a_n)$, or equivalently, if there exists $c
> 0$ such that $a_n \geq c  b_n$ for all $n$ sufficiently
large. Finally, we write $a_n = \Theta(b_n)$ if we have $a_n =
O(b_n)$ and $a_n = \Omega(b_n)$ at the same time.

\subsection{Organization of the Paper}
The paper is organized as follows: In Section \ref{sec:Model},  
we  give a formal model for the random pairwise key predistribution 
scheme of Chan et al., 
and introduce the induced random $K$-out graph. In particular, 
the main model $\mathbb{H \cap G}(n;K,p)$ considered in this paper, i.e., 
the intersection of a random $K$-out graph with
an Erd\H{o}s-R\'enyi graph, is introduced in Section \ref{subsec:intersection_random_graphs}.
The main result of the paper concerning the minimum node degree of
$\mathbb{H \cap G}(n;K,p)$ is presented in Section \ref{sec:Results}.
In Section \ref{sec:Comments}, we compare our results against the classical results
of Erd\H{o}s-R\'enyi and then against 
earlier results by Ya\u{g}an and Makowski \cite{YaganMakowskiPER} on the absence of isolated
nodes in $\mathbb{H \cap G}(n;K,p)$. Also in Section \ref{subsec:numerical}, we provide numerical
results in support of our analytical results. The proof of the main result is initiated 
in Section \ref{sec:Prelim} where we establish an important confining result that significantly eases the rest of the proof; there
we also establish some preliminary scaling results to be used throughout. The proof of our main result is outlined 
in Section \ref{sec:ProofTheoremNodeIsolation} and the necessary steps are established in Sections
\ref{sec:proof_prop_1} through \ref{sec:second3}.

\section{Model}
\label{sec:Model}

\subsection{The random pairwise key predistribution scheme}

Interest in the random pairwise key predistribution scheme of Chan
et al. \cite{ChanPerrigSong} stems from the following advantages
over the original Eschenauer - Gligor scheme: (i) Even if some nodes are captured, the
secrecy of the remaining nodes is {\em perfectly} preserved; (ii)
Unlike earlier schemes, this pairwise scheme enables both
node-to-node authentication and quorum-based revocation. See also 
\cite{YaganThesis} for a detailed 
comparison of these two classical key predistribution schemes.

We parametrize the pairwise key
distribution scheme by two positive integers $n$ and $K$ such that
$K < n$. There are $n$ nodes, labelled $i=1, \ldots , n$, with
unique ids ${\rm Id}_1, \ldots , {\rm Id}_n$. Write ${\cal V} =
\{ 1, \ldots, n \}$ and set ${\cal V}_{-i} = {\mathcal{V}}-\{i\}$ for
each $i=1, \ldots , n$. With node $i$, we associate a subset
$\Gamma_{n,i}(K)$ of $K$ nodes selected uniformly at {\em random} 
from ${\cal V}_{-i}$, We say that each of the nodes 
in $\Gamma_{n,i}(K)$ is paired to node $i$. 
Thus, for any subset $A \subseteq {\cal V}_{-i}$, we require
\begin{equation}
\bP{ \Gamma_{n,i}(K) = A } = \left \{
\begin{array}{ll}
{{n-1}\choose{K}}^{-1} & \mbox{if $|A|=K$} \\
              &                   \\
0             & \mbox{otherwise.} \\
\end{array}
\right .
\label{eq:defn_of_Gamma}
\end{equation}
Put differently, the selection of $\Gamma_{n,i}(K)$ is done {\em uniformly} amongst
all subsets of ${\cal V}_{-i}$ which are of size $K$ and we further assume that
rvs $\Gamma_{n,1}(K), \ldots , \Gamma_{n,n}(K)$ are
mutually independent.

Once this {\em offline} random pairing has been created, we
construct the key rings $\Sigma_{n,1}(K), \ldots , \Sigma_{n,n}(K)$,
one for each node, as follows: Assumed available is a collection of
$nK$ distinct cryptographic keys $\{ \omega_{i|\ell}, \ i=1,
\ldots , n ; \ \ell=1, \ldots , K \}$. Fix $i=1, \ldots , n$ and
let $\ell_{n,i}: \Gamma_{n,i}(K) \rightarrow \{ 1, \ldots , K \}$
denote a labeling of $\Gamma_{n,i}(K)$. For each node $j$ in
$\Gamma_{n,i}(K)$ paired to $i$, the cryptographic key
$\omega_{i|\ell_{n,i}(j)}$ is associated with $j$. For instance,
if the random set $\Gamma_{n,i}(K)$ is realized as $\{ j_1, \ldots ,
j_K \}$ with $1 \leq j_1 < \ldots < j_K \leq n $, then an obvious
labeling consists in $\ell_{n,i}(j_k) = k $ for each $k=1, \ldots , K$ 
so that key $\omega_{i|k}$ is associated with node $j_k$. Of course
other labelings are possible. Finally, with node $j$ paired to
node $i$, the pairwise key 
$ \omega^\star_{n,ij} 
= [ {\rm Id}_i | {\rm Id}_j | \omega_{i|\ell_{n,i}(j)} ] $ 
is constructed and inserted in the
memory modules of both nodes $i$ and $j$. The key
$\omega^\star_{n,ij}$ is assigned {\em exclusively} to the pair of
nodes $i$ and $j$, hence the terminology pairwise predistribution
scheme. The key ring $\Sigma_{n,i}(K)$ of node $i$ is the set
\begin{eqnarray}
\Sigma_{n,i}(K) = \left \{ \omega^\star_{n,ij}, \ j \in \Gamma_{n,i}(K) \right \}
\cup 
\left \{ \omega^\star_{n,ji}, \ 
\begin{array}{c}
j=1, \ldots ,n \\
i \in \Gamma_{n,j}(K) \\
\end{array}
\right \}
\nonumber
\end{eqnarray}

Two nodes $i$ and $j$, can secure an existing communication link
 if and only if
$\Sigma_{n,i}(K) \cap \Sigma_{n,j}(K) \neq \emptyset$
which holds if at least one
of the events $i \in \Gamma_{n,j}(K)$ or $j \in \Gamma_{n,i}(K)$ 
takes place. Namely, it is plain that
\[
[\Sigma_{n,i}(K) \cap \Sigma_{n,j}(K) \neq \emptyset] = [i \in \Gamma_{n,j}(K)] \cup [j \in \Gamma_{n,i}(K)]
\]
Both events can take place, in which case the memory
modules of node $i$ and $j$ both contain the distinct keys
$\omega^\star_{n,ij}$ and $\omega^\star_{n,ji}$. It is
plain by construction that this scheme supports {\em distributed}
node-to-node authentication.



\subsection{Random $K$-out graphs}

The pairwise key predistribution scheme naturally
gives rise to the following class of random graphs: With $n=2,3,
\ldots $ and positive integer $K < n$, we say that the distinct
nodes $i$ and $j$ are K-adjacent, written $i \sim_K j$, if and
only if they have at least one key in common in their key rings,
namely
\begin{equation}
i \sim_K j \quad \mbox{iff} 
\quad \Sigma_{n,i}(K) \cap \Sigma_{n,j}(K)
\neq \emptyset . \label{eq:Adjacency}
\end{equation}
Let $\mathbb{H}(n;K)$ denote the undirected random graph on the
vertex set $\{ 1, \ldots , n \}$ induced by the adjacency notion
(\ref{eq:Adjacency}). This ensures that edges in $\mathbb{H}(n;K)$ represent pairs of sensors 
that have at least one cryptographic key in common, and thus that can securely communicate
over an {\em existing} communication channel.
Let $\lambda_n (K)$ define the edge assignment probability in $\mathbb{H}(n;K)$; i.e., we have
\begin{eqnarray}
\bP{i \sim_{K} j } =  \lambda_n (K) \label{eq:edge_prob_key_graph}
\end{eqnarray}
for any distinct $i,j \in \mathcal{V}$. It is easy to check that
\begin{eqnarray}
\lambda_n (K) = 1- \bP{i \not \in \Gamma_{n,j}(K) ~\cap~ j \not \in \Gamma_{n,i}(K)} =1-\left(\frac{{{n-2} \choose {K}}}{{{n-1} \choose {K}}}\right)^2 = \frac{2K}{n-1}-\left(\frac{K}{n-1}\right)^2.
\label{eq:LinkAssignmentinH}
\end{eqnarray}

The random graph $\mathbb{H}(n;K)$ is known in the literature
on random graphs as the random $K$-out graph
\cite{Bollobas,JansonLuczakRucinski}, or random $K$-orientable graph \cite{FennerFrieze}. 
Those references adopt the following definition, which can easily be seen to be equivalent to the adjacency 
condition (\ref{eq:Adjacency}):
For each of the $n$ vertices assign exactly $K$ arcs to
$K$ distinct vertices that are selected
uniformly at random, and then ignore the orientation of the arcs. 
The directed version of this graph (i.e., with the orientation of the arcs preserved)
has also been studied; e.g., see the work by Philips et al. \cite{PhilipsTowsleyWolf}, who showed that
the {\em diameter} of the directed $K$-out graph concentrates
almost surely on two values.

\subsection{Intersection of random graphs}
\label{subsec:intersection_random_graphs}
As mentioned earlier,
we assume a simple wireless communication
model that consists of independent channels,
each of which can be either on or off. 
Thus, with $p$ in $(0,1)$, 
let $\{B_{ij}(p), 1 \leq i < j \leq n\}$ 
denote i.i.d. $\{0, 1\}$-valued rvs with
success probability $p$. The channel between nodes $i$ and $j$ is
available (resp.~up) with probability $p$ and unavailable (resp.~down) 
with the complementary probability $1-p$.

Distinct nodes $i$ and $j$ are said to be B-adjacent, written $i
\sim_{B} j$, if $B_{ij}(p) = 1$. B-adjacency defines
the standard Erd\H{o}s-R\'enyi (ER) graph $\mathbb{G}(n;p)$ on the vertex set $\{ 1,
\ldots , n \}$ \cite{Bollobas}. Obviously,
$
\bP{ i \sim_{B} j} = p.
$

The random graph model studied here is obtained 
by {\em intersecting} the random graphs induced by 
the pairwise key predistribution scheme, and by the on-off communication model,
respectively. Namely, we consider the intersection of 
$\mathbb{H}(n;K)$ with the
ER graph $\mathbb{G}(n;p)$. In this case, distinct nodes $i$
and $j$ are said to be adjacent, written $i \sim j$, if and only
they are both K-adjacent and B-adjacent, namely
\begin{equation}
i \sim j ~~\quad \mbox{iff} ~~~\quad
\Sigma_{n,i}(K) \cap \Sigma_{n,j}(K) \neq \emptyset ~~ \mbox{and} ~~
B_{ij}(p)=1.
\label{eq:Adjacency_Intersection}
\end{equation}
The resulting {\em undirected} random graph defined on the vertex
set $\{1, \ldots, n\}$ through this notion of adjacency is denoted
$\mathbb{H\cap G}(n;K,p)$. The relevance of  $\mathbb{H\cap G}(n;K,p)$ in the context of
secure WSNs is now clear. Two nodes that are connected by an edge in $\mathbb{H\cap G}(n;K,p)$ 
share at least one cryptographic key {\em and}
have a wireless link available to them, so that they can establish a {\em secure communication link}.

Throughout we assume the collections of rvs 
$\{ \Gamma_{n,1}(K), \ldots , \Gamma_{n,n}(K) \}$ 
and $\{B_{ij}(p), 1 \leq i < j \leq n\}$ to be independent, 
in which case the edge occurrence
probability in $\mathbb{H\cap G}(n;K,p)$ is given by
\begin{eqnarray}
\bP{i \sim j} = \bP{ i \sim_{K} j } \bP{ i \sim_{B} j }
= p \lambda_n (K) .
\label{eq:edge_prob_intersectioN_graph}
\end{eqnarray}

\section{The result}
\label{sec:Results}

Our main technical result is given next. 
To fix the terminology, we refer to any mapping $K: \mathbb{N}_0 \rightarrow \mathbb{N}_0$ as a {\em scaling} 
(for random $K$-out graphs) provided it satisfies the natural conditions
\begin{equation}
K_n < n  \qquad n=1,2, \ldots.
\label{eq:ScalingDefn}
\end{equation} 
Similarly, we let any mapping $p: \mathbb{N}_0 \rightarrow [0,1]$
define a scaling for Erd\H{o}s-R\'enyi graphs.

To lighten the notation we often group the parameters $K$ and $p$ into the ordered 
pair $\theta \equiv (K,p)$. Hence, a mapping 
$\theta: \mathbb{N}_0 \rightarrow \mathbb{N}_0 \times [0,1] $ 
defines a scaling for the intersection graph 
$\mathbb{H \cap G}(n;\theta)$ provided that the condition (\ref{eq:ScalingDefn}) holds.

\begin{theorem}
{\sl Consider scalings $K: \mathbb{N}_0 \rightarrow \mathbb{N}_0$
and $p: \mathbb{N}_0 \rightarrow [0,1]$ such that
$\lim_{n \to \infty} (n-2K_n) = \infty$ and $\limsup_{n \to \infty} p_n < 1$.
With the sequence $\gamma: \mathbb{N}_0 \rightarrow \mathbb{R}$
defined through
\begin{equation}
\label{eq:scalinglaw}
p_nK_n\left(1-\frac{\log(1-p_n)}{p_n}-\frac{K_n}{n-1}\right)= \log n + (k-1)\log \log n + \gamma_n 
\end{equation}we have
\begin{eqnarray}
\lim_{n \rightarrow \infty } \bP{ \textrm{Min~node~degree~of~} \atop
\mathbb{H}\cap\mathbb{G}(n;\theta_n)~\textrm{
is~no~less~than~} k} = \left \{
\begin{array}{ll}
0 & \mbox{\textrm{if}~ $\lim\limits_{n \to \infty}\gamma_n = -\infty$} \\
  &                      \\
1 & \mbox{\textrm{if} ~$\lim\limits_{n \to \infty}\gamma_n = +\infty$}. 
\end{array}
\right . \label{eq:OneLaw+NodeIsolation}
\end{eqnarray}
} \label{thm:nodedegree}
\end{theorem}

The proof of Theorem \ref{thm:nodedegree} passes through the
method of first and second moments \cite{JansonLuczakRucinski}, 
applied to the random variable counting the number
of nodes with degree $\ell$, with $\ell = 0,1, \ldots, k-1$. Although this technique is
standard in the literature, its application to the intersection graph $\mathbb{H \cap G}(n; \theta)$
is far from being straightforward due to intricate dependencies amongst the degrees of nodes.
The proof of Theorem \ref{thm:nodedegree} is given in Sections \ref{sec:ProofTheoremNodeIsolation} through \ref{sec:second3}.

The extra conditions enforced by Theorem \ref{thm:nodedegree} are
required for technical reasons; i.e., for the method of moments to be applied successfully to the aforementioned
count variables. However, we remark that these conditions are mild and do not preclude their application in realistic
WSN scenarios. First, the condition $\limsup_{n \to \infty} p_n < 1$ enforces that wireless communication channels
between nodes do not become available with probability one as $n$ gets large. The situation $\limsup_{n \to \infty} p_n =1$ 
that is not covered by our result is reminiscent of the {\em full visibility} case considered in \cite{YaganMakowskiPairwise}, and is not likely to hold in practice. In fact, as the number of nodes gets large, it may be expected that $p_n$ goes to zero due to interference associated with a large number of nodes communicating simultaneously. Second, the condition $\lim_{n \to \infty} (n-2K_n) = \infty$ will already follow if $2 K_n \leq c n$ for some $c<1$. Given that $2K_n$ is equal to the mean number of keys stored per sensor in the pairwise scheme \cite{YM_PerfEval}, this condition needs to hold in any practical WSN scenario due to limited memory and computational capability of the sensors. In fact, Di Pietro et al. \cite{DiPietro_Tissec} 
noted that key ring sizes on the order of
 $\log n$ are feasible for WSNs.
 
 We now present a simple corollary of Theorem \ref{thm:nodedegree}, that will help in comparing our main result with the classical results of
 Erd\H{o}s-R\'enyi \cite{erdos61conn}. 
 
 \begin{corollary}
 {\sl Consider scalings $K: \mathbb{N}_0 \rightarrow \mathbb{N}_0$
and $p: \mathbb{N}_0 \rightarrow [0,1]$ such that
$\lim_{n \to \infty} (n-2K_n) = \infty$ and $\limsup_{n \to \infty} p_n < 1$.
With the sequence $\gamma: \mathbb{N}_0 \rightarrow \mathbb{R}$
defined through
\begin{equation}
\label{eq:scalinglaw_cor}
\frac{p_nK_n}{n-1}\left(1-\frac{\log(1-p_n)}{p_n}-\frac{K_n}{n-1}\right)= \frac{\log n + (k-1)\log \log n + \gamma_n}{n}
\end{equation}
we have
\begin{eqnarray}
\lim_{n \rightarrow \infty } \bP{ \textrm{Min~node~degree~of~} \atop
\mathbb{H}\cap\mathbb{G}(n;\theta_n)~\textrm{
is~no~less~than~} k} = \left \{
\begin{array}{ll}
0 & \mbox{\textrm{if}~ $\lim\limits_{n \to \infty}\gamma_n = -\infty$} \\
  &                      \\
1 & \mbox{\textrm{if} ~$\lim\limits_{n \to \infty}\gamma_n = +\infty$}. 
\end{array}
\right . \label{eq:OneLaw+NodeIsolation_cor}
\end{eqnarray}
} \label{cor:nodedegree}
 \end{corollary}
\myproof Pick scalings $K: \mathbb{N}_0 \rightarrow \mathbb{N}_0$
and $p: \mathbb{N}_0 \rightarrow [0,1]$ such that
$\lim_{n \to \infty} (n-2K_n) = \infty$ and $\limsup_{n \to \infty} p_n < 1$. Define the 
sequence $\gamma: \mathbb{N}_0 \rightarrow \mathbb{R}$ through
(\ref{eq:scalinglaw}). For this scaling, we have
\begin{eqnarray}
\frac{p_nK_n}{n-1}\left(1-\frac{\log(1-p_n)}{p_n}-\frac{K_n}{n-1}\right) &=& \frac{\log n + (k-1)\log \log n + \gamma_n}{n-1}
\nonumber \\ \nonumber
&=& \frac{\log n + (k-1)\log \log n + \gamma_n+\frac{\log n + (k-1)\log \log n + \gamma_n}{n-1}}{n}
\\ \label{eq:cor_comparison}
&=&  \frac{\log n + (k-1)\log \log n + \gamma_n (1+o(1))+ o(1)}{n}.
\end{eqnarray} 
Comparing (\ref{eq:cor_comparison}) with (\ref{eq:scalinglaw_cor}), we get the desired result 
(\ref{eq:OneLaw+NodeIsolation_cor}) from (\ref{eq:OneLaw+NodeIsolation}) as we note that  
\begin{eqnarray}
\lim_{n \to \infty} \gamma_n &=& + \infty \qquad \textrm{if and only if} \qquad \lim_{n \to \infty} \left( \gamma_n (1+o(1))+ o(1) \right) = +\infty
\nonumber \\
\lim_{n \to \infty} \gamma_n &=& - \infty \qquad \textrm{if and only if} \qquad \lim_{n \to \infty} \left( \gamma_n (1+o(1))+ o(1) \right) = - \infty.
\nonumber
\end{eqnarray}
\hfill\fsquare 

\section{Comments and Discussion}
\label{sec:Comments}


\subsection{Comparison with Erd\H{o}s-R\'enyi Graphs}

For each $p$ in $[0,1]$ and $n=2,3, \ldots $, let $\mathbb{G}(n;p)$ 
denote the Erd\H{o}s-R\'enyi graph on the vertex set $\{ 1, \ldots , n\}$
with edge probability $p$. 
It is known that edge assignments are mutually independent in $\mathbb{G}(n;p)$,
whereas they are strongly correlated in $\mathbb{H}(n; K)$
in that they are {\em negatively associated} in the
sense of Joag-Dev and Proschan \cite{JoagdevProschan}; see
\cite{YaganMakowskiPER} for details.
Thus, $\mathbb{H}(n;K)$ cannot be equated with $\mathbb{G}(n;p)$ 
even when the parameters $p$ and $K$ are selected 
so that the edge assignment probabilities in these two graphs
coincide, say $\lambda(n;K) = p$. 
Therefore, $\mathbb{H \cap G}(n;\theta)$
cannot be equated with an ER graph either, and the results
obtained here are {\em not} mere consequences of
classical results for ER graphs.

However, some similarities do exist between $\mathbb{H\cap
G}(n;\theta)$ and ER graphs. We start by presenting the following well-known
zero-one law for $k$-connectivity in ER graphs \cite{erdos61conn}: 
For any scaling $p: \mathbb{N}_0
\rightarrow [0,1]$ satisfying
\begin{equation}
p_n =  \frac{ \log n + (k-1) \log \log n + \gamma_n}{n}
\label{eq:scaling_ER}
\end{equation}
for some $\gamma:\mathbb{N}_0 \to \mathbb{R}$, it holds that
\begin{eqnarray}
\lefteqn{\lim_{n \rightarrow \infty } 
\bP{ ~ \mathbb{G}(n;p_n)  \mbox{~is $k$-connected~} }} &&
 \nonumber \\
&=& \lim_{n \rightarrow \infty} \bP{ ~ \mathbb{G}(n;p_n) \mbox{~has min. node degree $\geq k$~}  } 
= \left \{
\begin{array}{ll}
0 & \mbox{if~ $\gamma_n \to - \infty$} \\
  &                 \\
1 & \mbox{if~ $\gamma_n \to + \infty$.}
\end{array}
\right .
\label{eq:k_con_for_ER}
\end{eqnarray}

We now compare this with our main result by means of Corollary \ref{cor:nodedegree}. 
Notice that the right-hand sides of the scalings (\ref{eq:scalinglaw_cor}) and (\ref{eq:scaling_ER}) are exactly the same, and so are the 
corresponding zero-one laws (\ref{eq:OneLaw+NodeIsolation_cor}) and (\ref{eq:k_con_for_ER}), respectively. 
In the case of the ER graph $\mathbb{G}(n;p_n)$, the left-hand side of (\ref{eq:scaling_ER}) corresponds to the edge probability 
$p_n$. We now explore how the left-hand side of (\ref{eq:scalinglaw_cor}) is related to the corresponding edge probability
$p_n \lambda_n(K_n)$ (viz. (\ref{eq:edge_prob_intersectioN_graph})) of the graph $\mathbb{H}\cap\mathbb{G}(n;\theta_n)$.
First, we recall (\ref{eq:LinkAssignmentinH})
and use the fact that $\log (1-p_n) \leq - p_n$ to get
\[
\frac{p_nK_n}{n-1}\left(1-\frac{\log(1-p_n)}{p_n}-\frac{K_n}{n-1}\right) \geq p_n\lambda_n(K_n).
\]
Hence, in ER graphs the threshold of $k$-connectivity,
and of minimum node degree being at least $k$, appears when the link probability
is compared against $(\log n + (k-1) \log \log n)/n$. In $\mathbb{H \cap G} (n;\theta_n)$,
our result shows that the threshold appears when a quantity that is always larger than
the link probability $p_n \lambda_n(K_n)$ is compared against $(\log n + (k-1) \log \log n)/n$.
This indicates that $\mathbb{H \cap G} (n;\theta_n)$ tends to exhibit the property that all nodes have at least
$k$ neighbors {\em easier} than ER graphs; i.e., this property can be ensured by a smaller
link probability between nodes (which leads to smaller average node degree).

The situation is more intricate if it holds that $\lim_{n \to \infty} p_n =0$, whence we 
have 
\[
\log(1-p_n) = - p_n - \frac {p_n^2}{2} (1+o(1)).
\]
This leads to
\begin{eqnarray}
\frac{p_nK_n}{n-1}\left(1-\frac{\log(1-p_n)}{p_n}-\frac{K_n}{n-1}\right) &=&
\frac{p_nK_n}{n-1}\left(2-\frac{K_n}{n-1}+\frac{p_n}{2}(1+o(1))\right) \qquad\qquad\qquad\qquad
\nonumber \\
&=&\frac{p_nK_n}{n-1}\left(2-\frac{K_n}{n-1}\right)\left(1+\frac{p_n}{2}\cdot\frac{1+o(1)}{2-\frac{K_n}{n-1}}\right)
\nonumber \\
&=& p_n \lambda_n(K_n) (1+\Theta(p_n))
\label{eq:threshold_with_poor_channels_1}
 \\
&=& p_n \lambda_n(K_n) (1+o(1)),
\label{eq:threshold_with_poor_channels}
\end{eqnarray}
where in (\ref{eq:threshold_with_poor_channels_1}), we used the fact that $1 \leq 2- \frac{K_n}{n-1} \leq 2$ since $K_n \leq n-1$.
Thus, in the practically relevant case when the wireless channels become weaker as 
$n$ gets large, the threshold for minimum node degree of $\mathbb{H \cap G} (n;\theta_n)$
to be at least $k$ appears when a quantity that is asymptotically equivalent to link probability 
is compared against $(\log n + (k-1) \log \log n)/n$; a situation that is reminiscent
of the ER graphs. A similar observation was made in
\cite{YaganMakowskiPER} for the threshold of $1$-connectivity and absence of isolated nodes.


Nevertheless, it is worth mentioning that even under $\lim_{n \to \infty} p_n =0$, the zero-one laws 
for the minimum node degree being at least $k$ in ER graphs and $\mathbb{H \cap G} (n;\theta_n)$ 
are {\em not} exactly analogous. This is because, the term $o(1)$ in (\ref{eq:threshold_with_poor_channels})
may change the behavior of the sequence $\gamma_n$ appearing in (\ref{eq:scalinglaw_cor}) as it is given by
\begin{eqnarray}
\nonumber
\gamma_n &=& n p_n \lambda_n(K_n) (1+\Theta(p_n)) - \log n - (k-1)\log \log n
 \\ \nonumber 
 &=& n p_n \lambda_n(K_n)  - \log n - (k-1)\log \log n + \Theta(n p_n^2 \lambda_n(K_n))
 \\ \nonumber
 &=& n p_n \lambda_n(K_n)  - \log n - (k-1)\log \log n + \Theta(K_n p_n^2 )
\end{eqnarray}
as we note that 
$\lambda_n(K_n) = \Theta(K_n /n)$.
It is now clear that, even under $\lim_{n \to \infty} p_n =0$, 
the two results, (\ref{eq:k_con_for_ER}) under (\ref{eq:scaling_ER}) and (\ref{eq:OneLaw+NodeIsolation_cor}) 
under (\ref{eq:scalinglaw_cor}), may be deemed
analogous if and only if $K_n p_n^2$ is bounded, i.e., $K_n p_n^2 = O(1)$. Combining, we can conclude
that for the two graphs, $\mathbb{G}(n;p_n)$ and $\mathbb{H \cap G} (n;K_n,p_n)$, to exhibit 
asymptotically the same behavior for the property that their minimum node degrees are at least $k$, the parameter 
scalings should satisfy
\[
p_n = o(1) \qquad \textrm{and} \qquad K_n p_n^2 = O(1).
\]

\subsection{Comparison with results by Ya\u{g}an and Makowski for $k=1$}
\label{subsec:CompYM}
We now compare our results with those by Ya\u{g}an and Makowski
\cite{YaganMakowskiPER} who established zero-one laws for 1-connectivity, and
for the absence of isolated nodes (i.e., absence of nodes with degree zero)
in $\mathbb{H \cap G}(n;\theta)$.
Here, we present their result in a slightly different form:
Consider scalings $K: \mathbb{N}_0 \rightarrow \mathbb{N}_0$
and $p: \mathbb{N}_0 \rightarrow (0,1)$ such that
\begin{equation}
p_n K_n \left(2-\frac{K_n }{n-1}\right) \left( \frac{1-\frac{\log(1-p_n)}{p_n}}{2} \right) \sim c  \log n, 
 \label{eq:scalinglaw_old}
\end{equation}
for some $c>0$. Assume also that $\lim_{n \to \infty}p_n=p^\star$
exists. Then, we have
\begin{eqnarray}
\lefteqn{ \lim_{n \rightarrow \infty } \bP{ \mathbb{H \cap
G}(n;\theta_n)~\mbox{is connected} } } &&
\nonumber \\
&=& \lim_{n \rightarrow \infty } \bP{\mathbb{H \cap
G}(n;\theta_n)~\mbox{contains~no~isolated~nodes } }
= \left \{
\begin{array}{ll}
0 & \mbox{if~ $c < 1$} \\
  &                      \\
1 & \mbox{if~ $c > 1$}.
\end{array}
\right . \label{eq:OneLaw}
\end{eqnarray}

To better compare this result with ours, we set $k=1$ and rewrite our scaling condition (\ref{eq:scalinglaw})
as
\begin{equation}
p_n K_n \left(2-\frac{K_n }{n-1}\right) \left( \frac{1-\frac{\log(1-p_n)}{p_n}-
\frac{K_n}{n-1}}{2-\frac{K_n}{n-1}} \right) = \log n + \gamma_n
\label{eq:scalinglaw_new}
\end{equation}
under which Theorem \ref{thm:nodedegree} gives
\begin{eqnarray}\nonumber
\lim_{n \rightarrow \infty } \bP{ \mathbb{H}\cap\mathbb{G}(n;\theta_n)~\atop {\textrm{
has~no~isolated~nodes~}}}
= \left \{
\begin{array}{ll}
0 & \mbox{\textrm{if}~ $\gamma_n \to -\infty$} \\
  &                      \\
1 & \mbox{\textrm{if} ~$\gamma_n \to +\infty$}. 
\end{array}
\right . 
\end{eqnarray}

We now argue how our result on absence of isolated nodes
constitutes an improvement on the result of \cite{YaganMakowskiPER}. The assumption that
limit $\lim_{n \to \infty} p_n = p^{\star}$ exists was the key in establishing (\ref{eq:OneLaw})
under (\ref{eq:scalinglaw_old}) and our results in this paper explains why. First, it is clear that if
$p^{\star}=0$, then
\[
\lim_{n \to \infty} \left( \frac{1-\frac{\log(1-p_n)}{p_n}}{2} \right) = 1 = \lim_{n \to \infty}
\left( \frac{1-\frac{\log(1-p_n)}{p_n}-
\frac{K_n}{n-1}}{2-\frac{K_n}{n-1}} \right)
\]
so that the left hand sides of (\ref{eq:scalinglaw_new}) and (\ref{eq:scalinglaw_old})
are asymptotically equivalent. Next, if $p^{\star}>0$, then it follows that $K_n = O(\log n)$
(see \cite{YaganMakowskiPER}) under (\ref{eq:scalinglaw_old}). This again yields the
asymptotical equivalence of the
 left hand sides of (\ref{eq:scalinglaw_new}) and (\ref{eq:scalinglaw_old}). Therefore, under
the assumption that $p_n$ has a limit, a scaling condition that is {\em equivalent} to (\ref{eq:scalinglaw_old}) 
is given by
\begin{eqnarray}
p_n K_n \left(2-\frac{K_n }{n-1}\right) \left( \frac{1-\frac{\log(1-p_n)}{p_n}-
\frac{K_n}{n-1}}{2-\frac{K_n}{n-1}} \right)  \sim c  \log n, 
 \label{eq:scalinglaw_old_2}
\end{eqnarray}
with the results (\ref{eq:OneLaw}) unchanged.

Comparing (\ref{eq:scalinglaw_new}) with (\ref{eq:scalinglaw_old_2}), 
we see that our absence of isolated nodes 
result is 
 more fine-grained than the one given in \cite{YaganMakowskiPER}.
 In a nutshell, the scaling condition
(\ref{eq:scalinglaw_old_2}) enforced in
\cite{YaganMakowskiPER}
requires a deviation of $\gamma_n = \pm \Omega(\log n)$ (from the threshold $\log n$)
 to get the zero-one law, whereas in our formulation (\ref{eq:scalinglaw_new}),
it suffices to have an unbounded deviation; e.g., even $\gamma_n = \pm \log \log \cdots \log n$ will do.
Put differently, we cover the case of $c=1$ in (\ref{eq:OneLaw}) under (\ref{eq:scalinglaw_old_2})
and show that $\mathbb{H \cap G} (n; \theta_n)$ could be
almost surely free of or not free of isolated nodes, depending on the limit
of ${\gamma_n}$; in fact, if  (\ref{eq:scalinglaw_old_2}) holds with 
$c>1$, we see from Theorem \ref{thm:nodedegree} that $\mathbb{H \cap G} (n; \theta_n)$ is
not only free of isolated nodes but also all of its nodes will have degree larger than $k$ 
for all $k=1,2, \ldots$.

\subsection{Numerical results and a conjecture}
\label{subsec:numerical}
We now present some numerical results to check the
validity of Theorem \ref{thm:nodedegree}, particularly in the non-asymptotic regime, i.e.,
when parameter values are set in accordance with real-world wireless sensor network scenarios.
In all experiments, 
we fix the number of nodes at $n=2000$. Then
for a given parameter pair $(K, p)$, we generate $200$
independent samples of the graph 
$\mathbb{H \cap G}(n;K,p)$ and count the number of times (out of a
possible 200) that the obtained graphs i) have minimum node degree 
no less than $k$
and ii) are $k$-connected, for $k=1, 2, \ldots$. Dividing the counts by $200$, we obtain the
(empirical) probabilities for the events of interest. 

\begin{figure*}[!t]
 \centering\subfigure[]
{
\includegraphics[totalheight=0.28\textheight,
width=0.5\textwidth]{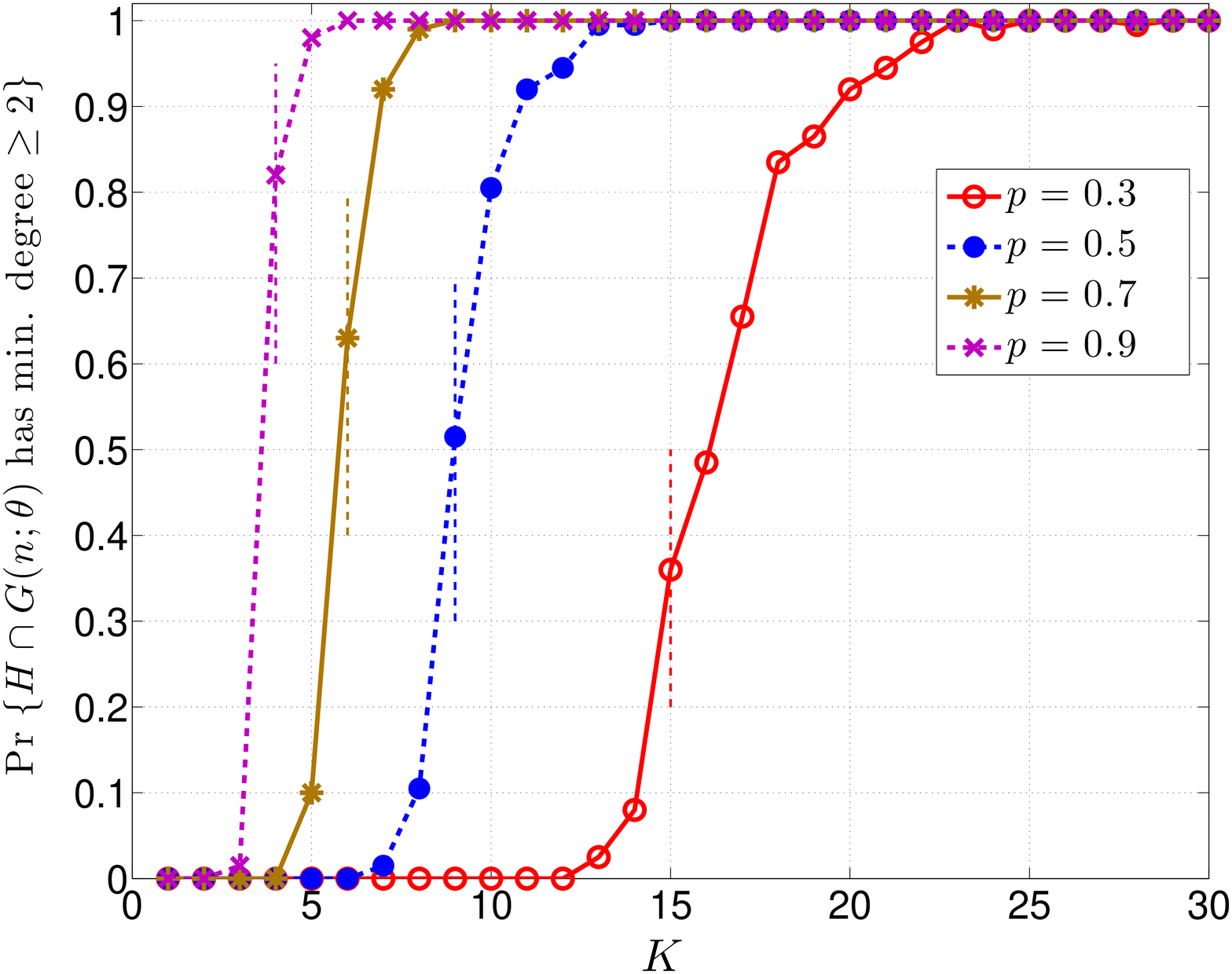} \label{fig:k_2_min} }
\subfigure[] {\hspace{-0.5cm}
\includegraphics[totalheight=0.28\textheight,
width=0.5\textwidth]{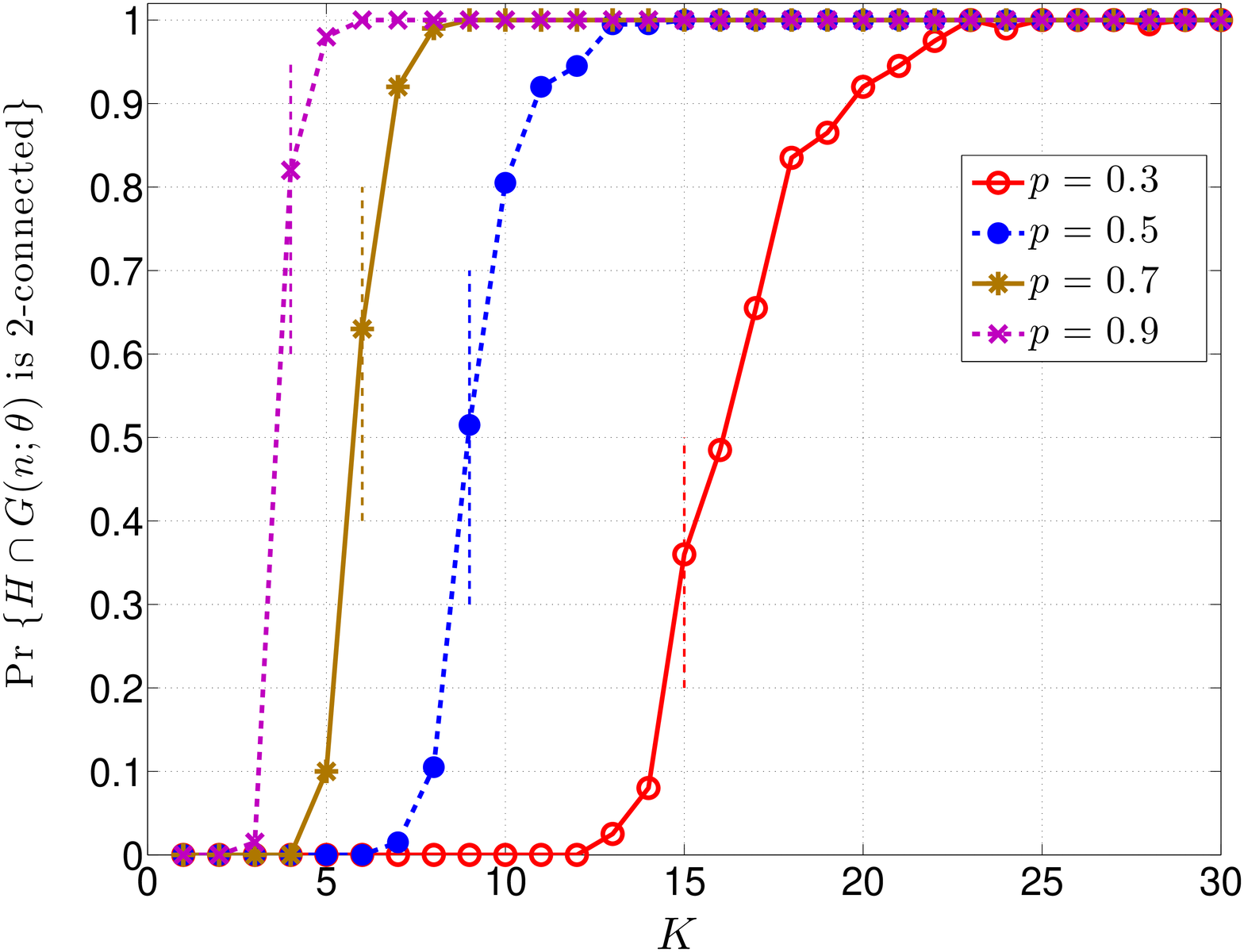}\label{fig:k_2_con}} \caption{\sl
{a) Probability that all nodes in
$\mathbb{H} \cap \mathbb{G}(n;K,p)$
         have degree at least 2 as a function of $K$ for
         $p=0.3$, $p=0.5$, $p=0.7$, and $p=0.9$
         with $n=2000$. b) Probability that
$\mathbb{H} \cap \mathbb{G}(n;K,p)$
         is 2-connected as a function of $K$ for
         $p=0.3$, $p=0.5$, $p=0.7$, and $p=0.9$
         with $n=2000$. The two figures being indistinguishable
         suggests that an analog of Theorem 
         \ref{thm:nodedegree} holds also for the
         property of $k$-connectivity. }  }
\end{figure*}

In Figure \ref{fig:k_2_min}, we depict the resulting empirical
probability that each node in $\mathbb{H \cap G}(n;K,p)$ has degree at least $2$
as a function of $K$ for various $p$ values.
For each $p$ value, we also show the critical
threshold of having minimum degree at least $2$ asserted by Theorem
\ref{thm:nodedegree} (viz. (\ref{eq:scalinglaw})) by a vertical dashed line. Namely,
the vertical dashed lines stand for the minimum integer value of
$K$ that satisfies
\begin{equation}
 p K \left(1-\frac{\log(1-p)}{p}-\frac{K}{n-1}\right) >
\log n + \log \log n \label{eq:threshold}
\end{equation}
Even with $n=2000$, we can observe the threshold 
behavior suggested by Theorem \ref{thm:nodedegree}; 
i.e., the probability that $\mathbb{H \cap G}(n;K,p)$ has minimum node degree at least $k$
transitions from {\em zero} to {\em one} as $K$ varies very slightly from a certain value.
Those $K$ values match well the vertical dashed lines suggested by Theorem \ref{thm:nodedegree},
leading to the conclusion that
numerical experiments are in good
agreement with our theoretical results.

\begin{figure*}[!h]
 \centering\subfigure[]
{
\includegraphics[totalheight=0.28\textheight,
width=0.5\textwidth]{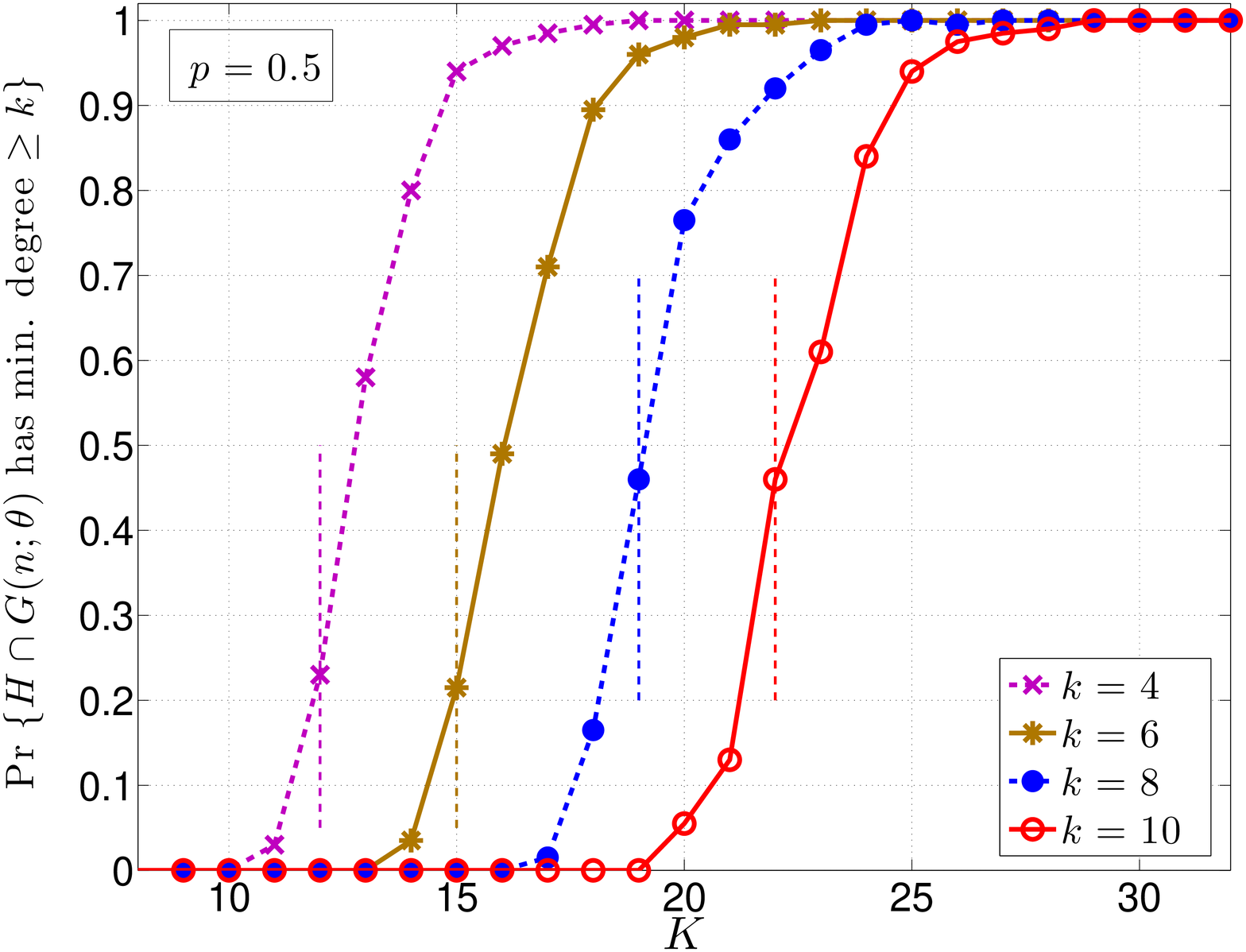} \label{fig:varying_k_min} }
\subfigure[] {\hspace{-0.5cm}
\includegraphics[totalheight=0.28\textheight,
width=0.5\textwidth]{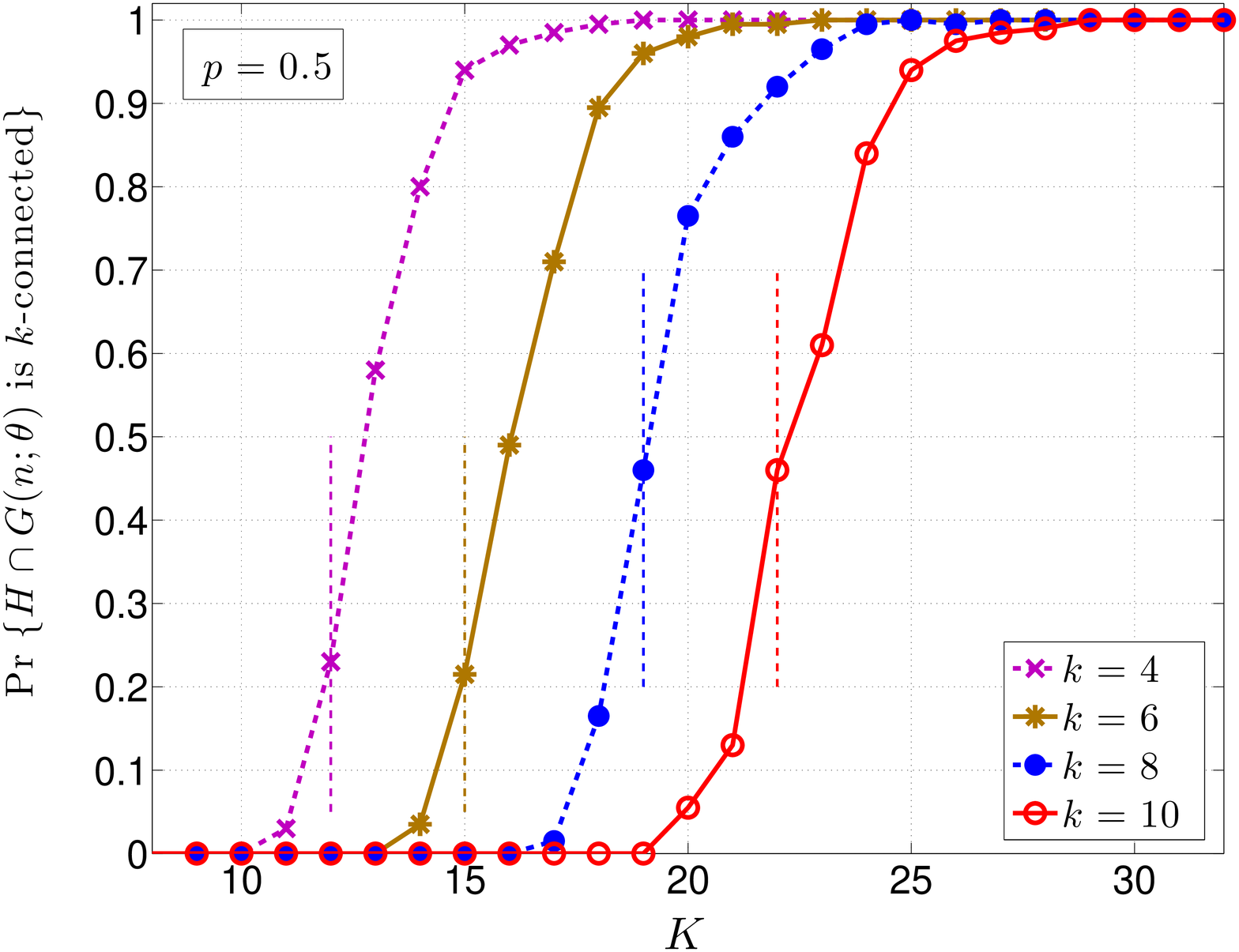}\label{fig:varying_k_con}} 
\subfigure[]{
\includegraphics[totalheight=0.28\textheight,
width=0.5\textwidth]{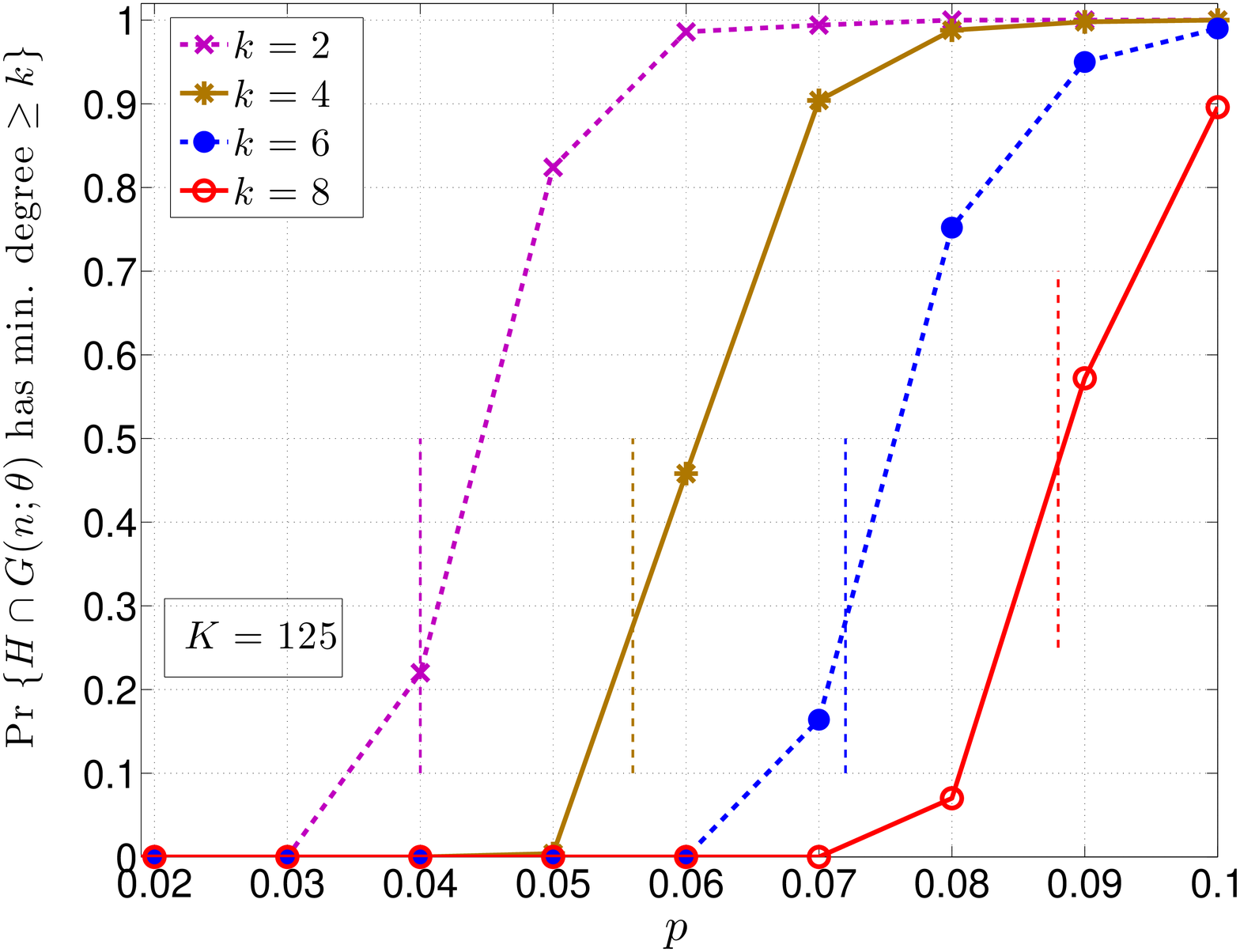} \label{fig:varying_p_min} }
\subfigure[] {\hspace{-0.5cm}
\includegraphics[totalheight=0.28\textheight,
width=0.5\textwidth]{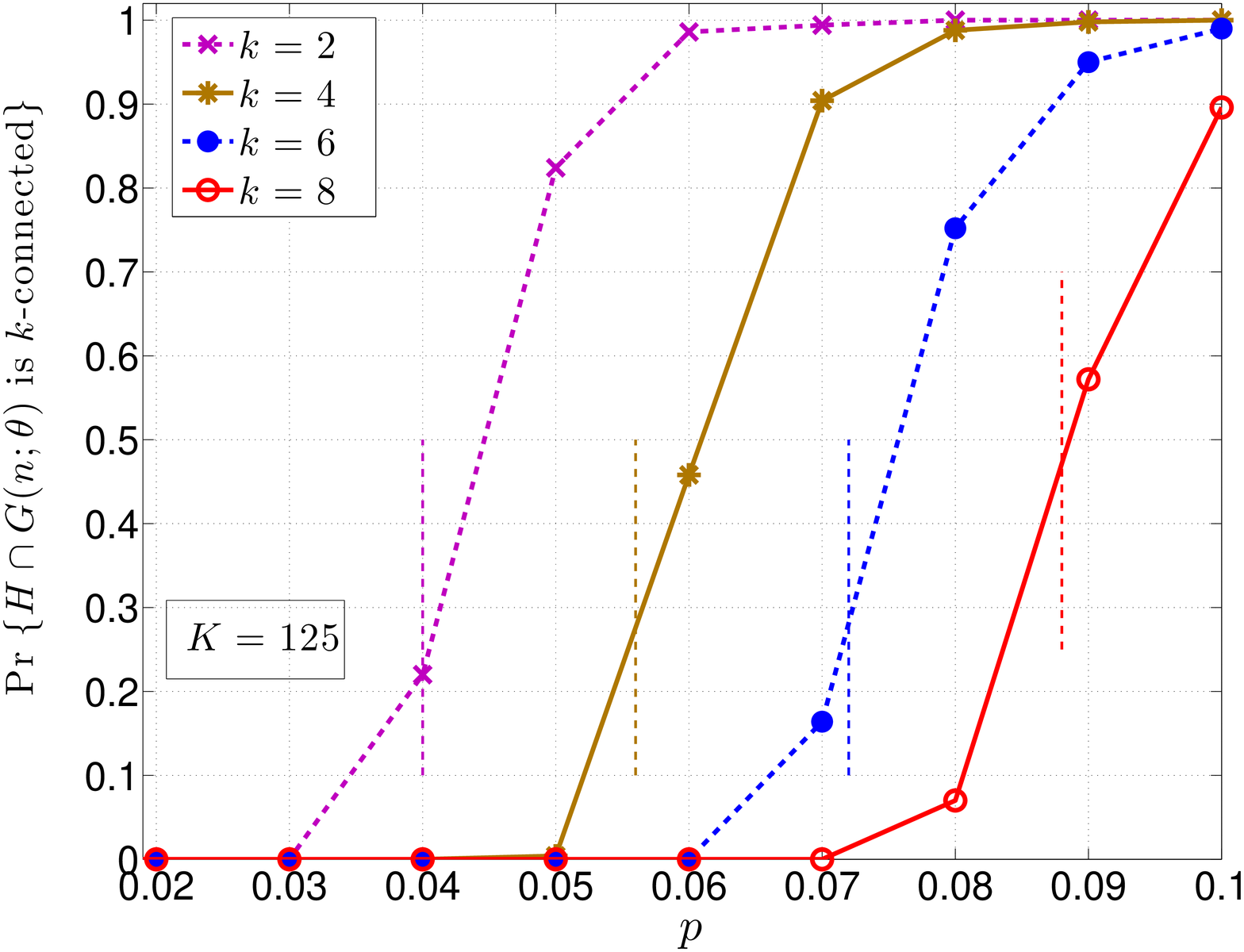}\label{fig:varying_p_con}} 
\caption{\sl
{a,b) With $n=2000$ and $p=0.5$, the probability that all nodes in
$\mathbb{H} \cap \mathbb{G}(n;K,p)$
         have degree at least $k$, and the probability that  
         $\mathbb{H} \cap \mathbb{G}(n;K,p)$
         is $k$-connected are plotted, respectively, as a function of $K$. 
         c,d) With $n=2000$ and $K=125$, the probability that all nodes in
$\mathbb{H} \cap \mathbb{G}(n;K,p)$
         have degree at least $k$, and the probability that
         $\mathbb{H} \cap \mathbb{G}(n;K,p)$
         is $k$-connected are plotted, respectively as a function of $p$.   }  }
\end{figure*}

Figure \ref{fig:k_2_con} is obtained in the same way with 
Figure \ref{fig:k_2_min}, this time for the probability 
that $\mathbb{H \cap G}(n;K,p)$ is $2$-connected.\footnote{The definition of 
$k$-connectivity given here coincides with the notion of $k$-vertex-connectivity used in the literature.
$k$-vertex-connectivity formally states that the graph will remain connected
despite the deletion of any $k-1$ vertices, and $k$-edge-connectivity is defined similarly for the deletion of edges. Since 
$k$-vertex-connectivity
implies $k$-edge-connectivity \cite{erdos61conn}, 
we say that a graph is simply $k$-connected (without referring to vertex-connectivity) 
to refer to the fact that it will remain connected despite the deletion of any $k-1$ nodes {\em or} edges.} 
It is clear 
that two figures show a strong similarity with curves corresponding to
each $p$ value being almost indistinguishable.
This raises the possibility that an analog of the zero-one law 
given in Theorem \ref{thm:nodedegree} holds also for the property of
 $k$-connectivity in $\mathbb{H}\cap \mathbb{G}(n;K,P)$. This would be reminiscent of
 several other random graph models from the literature where the two graph properties
 (min. node degree $\geq k$ and $k$-connectivity) shown to be asymptotically equivalent; e.g.,
 see ER graphs \cite{erdos61conn} (viz. (\ref{eq:k_con_for_ER})), random key graphs and their intersection with ER graphs
 \cite{RybarczykCoupling,ZhaoYaganGligor}, and random geometric graphs over a unit torus \cite{PenroseBook}.

To drive this point further, we have conducted an extensive simulation 
study and compared the empirical probabilities for the properties of 
minimum node degree is at least $k$, and $k$-connectivity in graph $\mathbb{H}\cap \mathbb{G}(n;K,P)$. 
Some of the results are reported in Figures \ref{fig:varying_k_min}-\ref{fig:varying_p_con}, 
and they strongly suggest the equivalence of these two properties in  $\mathbb{H}\cap \mathbb{G}(n;K,P)$
as well. This leads us to cast the following conjecture, which is the analog of Theorem \ref{thm:nodedegree}
for $k$-connectivity.
\begin{conjecture}
{\sl Consider scalings $K: \mathbb{N}_0 \rightarrow \mathbb{N}_0$
and $p: \mathbb{N}_0 \rightarrow [0,1]$ such that
$\lim_{n \to \infty} (n-2K_n) = \infty$ and $\limsup_{n \to \infty} p_n < 1$, and
a sequence $\gamma: \mathbb{N}_0 \rightarrow \mathbb{R}$
defined through (\ref{eq:scalinglaw}). Then, 
\begin{eqnarray}
\lim_{n \rightarrow \infty } \bP{ 
\mathbb{H}\cap\mathbb{G}(n;\theta_n)~\textrm{is $k$-connected}} = \left \{
\begin{array}{ll}
0 & \mbox{\textrm{if}~ $\gamma_n \to -\infty$} \\
  &                      \\
1 & \mbox{\textrm{if} ~$\gamma_n \to +\infty$}. 
\end{array}
\right . \label{eq:OneLaw+k_con}
\nonumber
\end{eqnarray}
} \label{conj:k_con}
\end{conjecture}

We close this section with a few comments on Conjecture \ref{conj:k_con}, before we start the proof of our main result in the next section. 
First, it is clear that if a graph has minimum node degree less than $k$, i.e., it has at least one vertex whose degree is less 
than or equal to $k-1$, then it will be {\em not} $k$-connected. This is because the graph can be made disconnected
by taking all the neighbors of the node with degree $\leq k-1$; i.e., by taking less than or equal to $k-1$ nodes. Therefore, 
Theorem \ref{thm:nodedegree} already establishes the zero-law of the Conjecture \ref{conj:k_con}. Namely, it is clear  
under the enforced assumptions on the scalings that
\[
\lim_{n \rightarrow \infty } \bP{ 
\mathbb{H}\cap\mathbb{G}(n;\theta_n)~\textrm{is $k$-connected}} = 0 \qquad 
\textrm{if} \quad \gamma_n \to -\infty.
\]
Therefore, it only remains to establish the one-law in Conjecture \ref{conj:k_con}. In view of Theorem \ref{thm:nodedegree}, 
this will follow if it is shown that
\begin{equation}
\lim_{n \rightarrow \infty } \bP{ 
\mathbb{H}\cap\mathbb{G}(n;\theta_n)~\textrm{is {\em not} $k$-connected} ~ \bigcap~ \atop
~~~  \mathbb{H}\cap\mathbb{G}(n;\theta_n)~\textrm{has min.~degree $\geq k$}} = 0 \qquad 
\textrm{if} \quad \gamma_n \to +\infty.
\label{eq:to_finish_conjecture}
\end{equation}
Exploring the validity of (\ref{eq:to_finish_conjecture}) is one of the main directions to be followed in the future work.

\section{Preliminaries}
\label{sec:Prelim}
Before we give a proof of Theorem \ref{thm:nodedegree},
we collect in this section some preliminary results that will be used throughout.

\subsection{A reduction step: Confining $\gamma_n$}
\label{subsec:confining}
A key step in proving Theorem \ref{thm:nodedegree} is to restrict the deviation function
$\gamma_n$ defined through (\ref{eq:scalinglaw}) to satisfy $\gamma_n = \pm o(\log n)$; i.e., that
\begin{equation}
\lim_{n \to \infty} \frac{\gamma_n}{\log n} = 0.
\label{eq:gamma_condition}
\end{equation}
Some useful consequences of (\ref{eq:gamma_condition}) are established in Section \ref{subsec:useful}.
In this section, we will show that (\ref{eq:gamma_condition}) can be assumed without loss of generality 
in establishing Theorem \ref{thm:nodedegree}. More precisely, we will show that 
\begin{eqnarray}
\textrm{Theorem \ref{thm:nodedegree} under $\gamma_n = \pm o(\ln n)$}
&\Rightarrow& 
\textrm{Theorem \ref{thm:nodedegree}} \hspace{2cm}
\label{eq:partb_with_extra_implies_partb}
\end{eqnarray}

First, we establish the fact that $\gamma_n$ defined through (\ref{eq:scalinglaw}) is monotone increasing
in both parameters $p_n$ and $K_n$.
\begin{proposition}
{\sl With $p$ in $(0,1)$ and a positive integer $K < n$, the function
\begin{equation}
\gamma_n = p K \left(1-\frac{\log(1-p)}{p}-\frac{K}{n-1}\right) - \log n - (k-1)\log \log n
\label{eq:gamma_without_scaling}
\end{equation}
}
is monotone increasing in $p$ and $K$.
\label{prop:gamma_monotone}
\end{proposition}

\myproof We first show that $\gamma_n$ is monotone increasing in $p$. Taking the derivative of 
(\ref{eq:gamma_without_scaling}) with respect to $p$, we get
\begin{eqnarray}
\frac{d}{dp} \gamma_n
&=& \frac{d}{dp} \left(pK -K \log(1-p) - p \frac{K^2}{n-1}\right) 
\\  \nonumber
&=& K + K \frac{1}{1-p} - \frac{K^2}{n-1} 
\\ \label{eq:inter_in_monotone}
&\geq& K + K \frac{1}{1-p} -K
\\ \nonumber
&\geq& 0,
\end{eqnarray}
where, in (\ref{eq:inter_in_monotone}) we used the fact that $K \leq n-1$.

Next, we show that $\gamma_n$ is monotone increasing in $K$ as well. To see this, take the 
derivative of (\ref{eq:gamma_without_scaling}) with respect to $K$ to get
\begin{eqnarray}
\frac{d}{dK} \gamma_n
&=& \frac{d}{dK} \left(pK -K \log(1-p) - p \frac{K^2}{n-1}\right) 
\\  \nonumber
&=& p -  \log(1-p) - 2 \frac{Kp}{n-1} 
\\ \label{eq:inter2_in_monotone}
&\geq& p -\log(1-p) -2p
\\ \label{eq:inter3_in_monotone}
&\geq& 0,
\end{eqnarray}
where in (\ref{eq:inter2_in_monotone}) and (\ref{eq:inter3_in_monotone}), 
we used the facts that $K \leq n-1$ and $\log (1-p) \leq - p$, respectively.
\myendpf

Recall that any mapping $(K, p): \mathbb{N}_0 \to \mathbb{N}_0 \times [0,1]$ defines a scaling
provided that the condition (\ref{eq:ScalingDefn}) is satisfied. We now introduce the notion
of an {\em admissible} scaling. 

\begin{definition}
{\sl A mapping $(K, p): \mathbb{N}_0 \to \mathbb{N}_0 \times [0,1]$ is said to be an admissible scaling
if (\ref{eq:ScalingDefn}) holds, and the sequence $\gamma: \mathbb{N}_0 \rightarrow \mathbb{R}$
defined through (\ref{eq:scalinglaw}) satisfies (\ref{eq:gamma_condition}).
}
\end{definition}
The relevance of the notion of admissibility flows from the
following two results.

\begin{proposition}
{\sl Consider a scaling $(K, p): \mathbb{N}_0 \to \mathbb{N}_0 \times [0,1]$
such that $\lim_{n \to \infty} (n-2K_n) = \infty$, $\limsup_{n \to \infty} p_n < 1$,
and the sequence $\gamma: \mathbb{N}_0 \rightarrow \mathbb{R}$ defined through
(\ref{eq:scalinglaw}) satisfying
\[
\lim_{n \to \infty} \gamma_n = \infty.
\] 
Then, there always
exists an admissible scaling $(\tilde{K}, \tilde{p}): \mathbb{N}_0 \to \mathbb{N}_0 \times [0,1]$
with $\lim_{n \to \infty} (n-2\tilde{K}_n) = \infty$, $\limsup_{n \to \infty} \tilde{p}_n < 1$ and such that
\begin{equation}
 \tilde{K}_n \leq K_n \quad {\rm and} \quad  \tilde{p}_n \leq p_n, \quad
n=1,2, \ldots \label{eq:ComparingAlpha+A}
\end{equation}
whose deviation function $\tilde{\gamma} : \mathbb{N}_0 \rightarrow
\mathbb{R}$ defined through
\begin{equation}
\tilde{\gamma}_n = \tilde{p}_n \tilde{K}_n \left( 1-\frac{\log(1-\tilde{p}_n)}{\tilde{p}_n}-
\frac{\tilde{K}_n}{n-1}  \right) - \log n - (k-1)\log \log n
\label{eq:gamma_n_for_admissible}
\end{equation}
satisfies both conditions
\begin{equation}
\lim_{n \rightarrow \infty} \tilde \gamma_n = \infty
\label{eq:ComparingAlpha+B}
\end{equation}
and
\begin{equation}
\tilde \gamma_n = o(\log n). \label{eq:ComparingAlpha+C}
\end{equation}
\label{prop:admissible1}
}
\end{proposition}
\myproof Under the enforced assumptions on the scaling $(K, p): \mathbb{N}_0 \to \mathbb{N}_0 \times [0,1]$ and
the deviation sequence $\gamma_n$ associated with it,
pick $\tilde{K}_n = K_n$, $\tilde{\gamma}_n = \min\{\log \log n, \gamma_n \}$ for each $n=1,2, \ldots$, and define the sequence $\tilde{p}_n$
through
\begin{equation}
\tilde{\gamma}_n = \tilde{p}_n {K}_n \left( 1-\frac{\log(1-\tilde{p}_n)}{\tilde{p}_n}-
\frac{{K}_n}{n-1}  \right) - \log n - (k-1)\log \log n
\label{eq:intermediary_just_needed}
\end{equation}
Note that since $\tilde{\gamma}_n$ is monotone increasing in $\tilde{p}_n$ (see Proposition \ref{prop:gamma_monotone}), the relation (\ref{eq:intermediary_just_needed}) will uniquely define $\tilde{p}_n$.
Since $\tilde{\gamma}_n \leq \gamma_n$ by construction, we have $\tilde{p}_n \leq p_n$ in view of 
of the fact that deviation sequences are monotone increasing in $p$.
Thus, the pair $\tilde{K}_n, \tilde{p}_n$ satisfies (\ref{eq:ComparingAlpha+A}). It is also plain from $\tilde{\gamma}_n = \min\{\log \log n, \gamma_n \}$ and the fact that $\lim_{n \to \infty} \gamma_n = \infty$, that we have (\ref{eq:ComparingAlpha+B}) and (\ref{eq:ComparingAlpha+C}). Finally, it is clear that $\lim_{n \to \infty} (n-2\tilde{K}_n) = \infty$ (since $\tilde{K}_n = K_n$) and $\limsup_{n \to \infty} \tilde{p}_n < 1$ since $\tilde{p}_n \leq p_n$.
\myendpf

The next result is an analog of Proposition \ref{prop:admissible1} for the case $\lim_{n \to \infty}\gamma_n =-\infty$.
\begin{proposition}
{\sl Consider a scaling $(K, p): \mathbb{N}_0 \to \mathbb{N}_0 \times [0,1]$
such that $\lim_{n \to \infty} (n-2K_n) = \infty$, $\limsup_{n \to \infty} p_n < 1$,
and the sequence $\gamma: \mathbb{N}_0 \rightarrow \mathbb{R}$ defined through
(\ref{eq:scalinglaw}) satisfying
\[
\lim_{n \to \infty} \gamma_n = - \infty.
\] 
Then, there always
exists an admissible scaling $(\tilde{K}, \tilde{p}): \mathbb{N}_0 \to \mathbb{N}_0 \times [0,1]$
with $\lim_{n \to \infty} (n-2\tilde{K}_n) = \infty$, $\limsup_{n \to \infty} \tilde{p}_n < 1$ and such that
\begin{equation}
 \tilde{K}_n \geq K_n \quad {\rm and} \quad  \tilde{p}_n \geq p_n, \quad
n=1,2, \ldots \label{eq:ComparingAlpha+A_2}
\end{equation}
whose deviation function $\tilde{\gamma} : \mathbb{N}_0 \rightarrow
\mathbb{R}$ defined through
(\ref{eq:gamma_n_for_admissible})
satisfies both conditions
\begin{equation}
\lim_{n \rightarrow \infty} \tilde \gamma_n = - \infty
\label{eq:ComparingAlpha+B_2}
\end{equation}
and
\begin{equation}
\tilde \gamma_n = - o(\log n). \label{eq:ComparingAlpha+C_2}
\end{equation}
\label{prop:admissible2}
}
\end{proposition}

\myproof The proof of Proposition \ref{prop:admissible2} is a bit more tricky than that of 
Proposition \ref{prop:admissible1}. This time, we start by setting $\tilde{\gamma}_n = \max\{\gamma_n, -\log \log n\}$
under the enforced assumptions on the scalings 
$K_n$, $p_n$ and the associated deviation sequence $\gamma_n$
defined through (\ref{eq:scalinglaw}). It is plain that we have (\ref{eq:ComparingAlpha+B_2}) and (\ref{eq:ComparingAlpha+C_2}). Thus, we only need to find scalings $\tilde{p}_n$ and $\tilde{K}_n$ that satisfy
\begin{equation}
\tilde{\gamma}_n = \tilde{p}_n \tilde{K}_n \left( 1-\frac{\log(1-\tilde{p}_n)}{\tilde{p}_n}-
\frac{\tilde{K}_n}{n-1}  \right) - \log n - (k-1)\log \log n
\end{equation}
together with (\ref{eq:ComparingAlpha+A_2}), $\lim_{n \to \infty} (n-2\tilde{K}_n) = \infty$, 
and $\limsup_{n \to \infty} \tilde{p}_n < 1$. Since the deviation
sequence is monotone increasing (viz. Proposition \ref{prop:gamma_monotone}), and we have 
$\tilde{\gamma}_n \geq \gamma_n$, we can attempt to construct 
the scalings $\tilde{p}_n$ and $\tilde{K}_n$ as in the proof of Proposition \ref{prop:admissible1}. Namely, set $\tilde{K}_n = K_n$, choose
the sequence $\tilde{p}_n=\tilde{p}_n^{\star}$
that satisfies
\begin{equation}
\tilde{\gamma}_n = \tilde{p}_n^{\star} {K}_n \left( 1-\frac{\log(1-\tilde{p}_n^{\star})}{\tilde{p}_n^{\star}}-
\frac{{K}_n}{n-1}  \right) - \log n - (k-1)\log \log n
\label{eq:to_satisfy_tilde}
\end{equation}
It is plain that we
have $\tilde{p}_n^{\star} \geq p_n$ since $\tilde{\gamma}_n\geq \gamma_n$ .
If it holds that $\limsup_{n \to \infty} \tilde{p}_n^{\star} < 1$,
then we are done by choosing $\tilde{p}_n = \tilde{p}_n^{\star}$ and $\tilde{K}_n = K_n$. 

On the other hand, if
(\ref{eq:to_satisfy_tilde}) is satisfied with  $\limsup_{n \to \infty} \tilde{p}_n^{\star} = 1$, then we 
set 
\begin{equation}
\tilde{p}_n = \min \left \{ \max \{p_n, 0.5 \},\tilde{p}_n^{\star} \right\}, \qquad n=1,2, \ldots
\label{eq:novel_setting}
\end{equation}
so that $\tilde{p}_n \geq p_n$ and 
$\limsup_{n \to \infty} \tilde{p}_n < 1$ hold. Then pick a positive real number $\tilde{K}_n^{\star} \leq n-1$ such that it satisfies
\begin{equation}
\tilde{\gamma}_n = \tilde{p}_n \tilde{K}_n^{\star} \left( 1-\frac{\log(1-\tilde{p}_n)}{\tilde{p}_n}-
\frac{\tilde{K}_n^{\star}}{n-1}  \right) - \log n - (k-1)\log \log n,
\label{eq:to_satisfy_tilde2}
\end{equation}
and set $\tilde{K}_n = \lceil \tilde{K}_n^{\star} \rceil$. Note that $\tilde{K}_n^{\star}$ is 
uniquely defined from (\ref{eq:to_satisfy_tilde2}) since $\tilde{\gamma}_n$ is monotone increasing in 
$\tilde{K}_n^{\star}$ (see Proposition \ref{prop:gamma_monotone}). We will first show that the deviation sequence
associated with the pair $(\tilde{p}_n, \tilde{K}_n)$ is the same with that associated with $(\tilde{p}_n, \tilde{K}^{\star}_n)$
within an additive constant. Namely, with $\tilde{\gamma}_n'$ defined through
\[
\tilde{\gamma}_n' = \tilde{p}_n \tilde{K}_n \left( 1-\frac{\log(1-\tilde{p}_n)}{\tilde{p}_n}-
\frac{\tilde{K}_n}{n-1}  \right) - \log n - (k-1)\log \log n
\]
we will show that
\begin{equation}
\tilde{\gamma}_n'  = \tilde{\gamma}_n + O(1).
\label{eq:gamma_n_prime_equality}
\end{equation}
This will ensure that (\ref{eq:ComparingAlpha+B_2}) and (\ref{eq:ComparingAlpha+C_2}) are still in effect with
$\tilde{\gamma}_n'$ replaced by $\tilde{\gamma}_n$. In order to obtain (\ref{eq:gamma_n_prime_equality}), we note
that $\tilde{K}_n < \tilde{K}_n^{\star} +1$ and write
\begin{eqnarray}
\tilde{\gamma}_n' - \tilde{\gamma}_n \leq \tilde{p}_n - \log(1-\tilde{p}_n) - \tilde{p}_n \frac{2\tilde{K}_n^{\star} +1}{n-1}
\leq \tilde{p}_n - \log(1-\tilde{p}_n) = O(1),
\end{eqnarray}
where in the last step we used the fact that $\limsup_{n \to \infty} \tilde{p}_n < 1$.

Now, with $\tilde{K}_n = \lceil \tilde{K}_n^{\star} \rceil$ where $\tilde{K}_n^{\star} $ is defined in (\ref{eq:to_satisfy_tilde2}), we have to show that $\tilde{K}_n  \geq K_n$ and that $\lim_{n \to \infty} (n-2\tilde{K}_n) = \infty$.
First, since $\tilde{p}_n \leq \tilde{p}_n^{\star}$, we must have $\tilde{K}_n^{\star} \geq K_n$ so that the pair
($\tilde{p}_n, \tilde{K}_n^{\star}$) leads to same deviation sequence $\tilde{\gamma}_n$ with the pair 
($\tilde{p}_n^{\star}, {K}_n$); this is plain from Proposition \ref{prop:gamma_monotone}. This establishes
$\tilde{K}_n \geq K_n$ and the only condition which is yet to be shown is that $\lim_{n \to \infty} (n-2\tilde{K}_n) = \infty$. 
To see this,
first note from (\ref{eq:novel_setting}) that $\tilde{p}_n \geq 0.5$ for all $n$ sufficiently large
since we have $\limsup_{n \to \infty} \tilde{p}_n^{\star} = 1$. 
Then, from (\ref{eq:to_satisfy_tilde2}) we get 
\begin{eqnarray}
\tilde{\gamma}_n + \log n + (k-1)\log \log n &=& \tilde{p}_n \tilde{K}_n^{\star} \left( 1-\frac{\log(1-\tilde{p}_n)}{\tilde{p}_n}
- \frac{\tilde{K}_n^{\star}}{n-1}  \right)
\nonumber \\
 &\geq& \tilde{p}_n\tilde{K}_n^{\star}\left(2-  \frac{\tilde{K}_n^{\star}}{n-1}\right)   
 \nonumber \\
 & \geq&  \tilde{p}_n \tilde{K}_n^{\star} 
  \nonumber \\
 & \geq& 0.5 \tilde{K}_n^{\star}
 \end{eqnarray}
for all $n$ sufficiently large. Thus, in view of $\lim_{n \to \infty} \tilde{\gamma_n} = -\infty$, we conclude 
from the last inequality that $\tilde{K}_n^{\star} = O(\log n)$. Since 
$\tilde{K}_n^{\star} \leq \tilde{K}_n < \tilde{K}_n^{\star}+1$, this also ensures that
$\tilde{K}_n = O(\log n)$. The desired condition 
$\lim_{n \to \infty} (n-2\tilde{K}_n) = \infty$ is now established and this concludes the proof of Proposition \ref{prop:admissible2}.
\myendpf

Propositions \ref{prop:admissible1} and \ref{prop:admissible2} will pave the way in establishing the desired
reduction step (\ref{eq:partb_with_extra_implies_partb}) through the following coupling argument. In a nutshell,
the following result shows that the probability $\bP{ \textrm{Min~node~degree~of~} 
\mathbb{H}\cap\mathbb{G}(n;K_n,p_n)~\textrm{
is~no~less~than~} k}$ is monotone increasing in $K_n$ and $p_n$.
\begin{proposition}
{\sl Consider a scaling $(K, p): \mathbb{N}_0 \to \mathbb{N}_0 \times [0,1]$. Then for any
scaling $(\tilde{K}, \tilde{p}): \mathbb{N}_0 \to \mathbb{N}_0 \times [0,1]$ such that
(\ref{eq:ComparingAlpha+A_2})
holds,
we have
\begin{equation}
\bP{ \textrm{Min~node~degree~of~} \atop
\mathbb{H}\cap\mathbb{G}(n;K_n,p_n)~\textrm{
is~no~less~than~} k} \leq \bP{ \textrm{Min~node~degree~of~} \atop
\mathbb{H}\cap\mathbb{G}(n;\tilde{K}_n,\tilde{p}_n)~\textrm{
is~no~less~than~} k}
\label{eq:coupling_OY_1}
\end{equation}
Similarly, 
 for any
scaling $(\tilde{K}, \tilde{p}): \mathbb{N}_0 \to \mathbb{N}_0 \times [0,1]$ such that
(\ref{eq:ComparingAlpha+A})
holds,
we have
\begin{equation}
\bP{ \textrm{Min~node~degree~of~} \atop
\mathbb{H}\cap\mathbb{G}(n;K_n,p_n)~\textrm{
is~no~less~than~} k} \geq \bP{ \textrm{Min~node~degree~of~} \atop
\mathbb{H}\cap\mathbb{G}(n;\tilde{K}_n,\tilde{p}_n)~\textrm{
is~no~less~than~} k}
\label{eq:coupling_OY_2}
\end{equation}
}
\end{proposition}
\myproof It is plain that it suffices to establish only one of the desired results 
(\ref{eq:coupling_OY_1}) or (\ref{eq:coupling_OY_2}) 
under (\ref{eq:ComparingAlpha+A_2}) or (\ref{eq:ComparingAlpha+A}), respectively.
We will establish (\ref{eq:coupling_OY_1}) under (\ref{eq:ComparingAlpha+A_2})
by showing the existence of a coupling such that
$\mathbb{H}\cap\mathbb{G}(n;K_n,p_n)$ is a spanning subgraph of 
$\mathbb{H}\cap\mathbb{G}(n;\tilde{K}_n,\tilde{p}_n)$.
In this case, it is easy to conclude that (e.g., see the work by Rybarczyk \cite[pp. 7]{RybarczykCoupling})
\begin{eqnarray}
\mathbb{P}[\mathbb{H}\cap\mathbb{G}(n;K_n,p_n) \textrm{ has property }\mathcal{P}]
  \leq  \mathbb{P}[\mathbb{H}\cap\mathbb{G}(n;\tilde{K}_n, \tilde{p}_n) \textrm{ has property }\mathcal{P}] .
  \label{graph_g_a_g_b}
\end{eqnarray}
for any monotone increasing\footnote{A graph property
is called monotone increasing if it holds under the addition of
edges in a graph. \label{fnote:gr_prop}} graph property $\mathcal{P}$.
It is straightforward to see that the property that the minimum
node degree is no less than $k$ is monotone increasing
(and so is the property of $k$-connectivity).

We now show that, if (\ref{eq:ComparingAlpha+A_2}) holds, i.e., 
if $\tilde{K}_n \geq K_n$ and $\tilde{p}_n \geq p_n$, then
\begin{equation}
\mathbb{H}\cap\mathbb{G}(n;K_n,p_n) \subseteq \mathbb{H}\cap\mathbb{G}(n;\tilde{K}_n, \tilde{p}_n).
\label{eq:coupling_final_desired}
\end{equation}
The required coupling argument will be completed in two steps owing to the independence
of $\mathbb{H}(n;K_n)$ and $\mathbb{G}(n;p_n)$ in the construction of the intersection graph
$\mathbb{H}\cap\mathbb{G}(n;K_n,p_n)$. First, we argue that if $\tilde{p}_n \geq p_n$, then there exists
a coupling that establishes
\begin{equation}
\mathbb{G}(n;p_n) \subseteq \mathbb{G}(n; \tilde{p}_n) 
\label{eq:coupling_to_final_1}
\end{equation}
We use the same arguments as in \cite{ZhaoYaganGligor}. Pick independent 
Erd\H{o}s-R\'enyi graphs $\mathbb{G}(n, {{p}_n/ \tilde{p}_n})$ and
$\mathbb{G}(n, \tilde{p}_n)$ on the same vertex set; note that 
we can construct $\mathbb{G}(n, {{p}_n/ \tilde{p}_n})$
with link probability ${p}_n/ \tilde{p}_n$ since ${p}_n/ \tilde{p}_n \leq 1$
under the enforced assumptions.
It is plain that the intersection $\mathbb{G}(n, {{p}_n/ \tilde{p}_n}) \cap \mathbb{G}(n, \tilde{p}_n)$
will still be an Erd\H{o}s-R\'enyi graph (due to independence) with edge probability 
given by $\tilde{p}_n \cdot \frac{{p}_n}{\tilde{p}_n} = {p}_n$. In other words,
we have $\mathbb{G}(n, {{p}_n/ \tilde{p}_n}) \cap \mathbb{G}(n, \tilde{p}_n) =_{st} \mathbb{G}(n, {p}_n)$.
Consequently, under this coupling, $\mathbb{G}(n,{{p}_n\iffalse_{on}\fi})$ is a
spanning subgraph of $\mathbb{G}(n, \tilde{p}_n)$. 

Next, we provide a coupling argument that shows that
if $\tilde{K}_n \geq K_n$, then 
\begin{equation}
\mathbb{H}(n;K_n) \subseteq \mathbb{H}(n; \tilde{K}_n) 
\label{eq:coupling_to_final_2}
\end{equation}
Recall that $\mathbb{H}(n; \tilde{K}_n)$ is constructed as follows: assign each node
exactly $\tilde{K}_n$
arcs towards
$\tilde{K}_n$ distinct vertices that are selected
uniformly at random, and then ignore the orientation of the arcs. An equivalent way of
generating $\mathbb{H}(n; \tilde{K}_n)$ is as follows. In the first round, for each node assign
$K_n$ arcs towards $K_n$ vertices that are selected uniformly at random, and then ignore the orientation 
of the arcs. At this point, we have constructed an instantiation of $\mathbb{H}(n; {K}_n)$. Next,
to each node assign $\tilde{K}_n - K_n$ additional arcs towards $\tilde{K}_n - K_n$ distinct nodes which are
randomly selected from among all nodes that were not picked in the first round. Namely, for each node
this second round of selection
will be made uniformly at random among the set of $n-1-K_n$ nodes that were not picked in the previous
round. Finally, by ignoring the orientation of the arcs assigned in the second round, we obtain $\mathbb{H}(n; \tilde{K}_n)$.
It is now plain that we have (\ref{eq:coupling_to_final_2}) whenever $\tilde{K}_n \geq K_n$. 

The desired result (\ref{eq:coupling_final_desired}) follows immediately from (\ref{eq:coupling_to_final_1}) and (\ref{eq:coupling_to_final_2}) by independence. \myendpf

We can now establish (\ref{eq:partb_with_extra_implies_partb})
and reduce the proof of Theorem \ref{thm:nodedegree} to admissible scalings. Suppose that 
Theorem \ref{thm:nodedegree} is proved under the additional condition that the scaling
$(K, p): \mathbb{N}_0 \rightarrow \mathbb{N}_0 \times 
 [0,1]$ is admissible; i.e., that the associated deviation sequence
$\gamma_n$ defined through (\ref{eq:scalinglaw}) satisfies $\gamma_n = \pm o (\log n)$. 
This result is stated below as Proposition \ref{prop:nodedegree} for convenience.

Assume that Proposition \ref{prop:nodedegree} holds and 
pick any scaling $(K, p): \mathbb{N}_0 \rightarrow \mathbb{N}_0 \times  [0,1]$ such that 
$\lim_{n \to \infty} (n-2K_n) = \infty$, $\limsup_{n \to \infty} p_n < 1$.  If the deviation sequence
$\gamma_n$ defined through (\ref{eq:scalinglaw}) satisfies $\lim_{n \to \infty} \gamma_n = \infty$,
then we know from Proposition \ref{prop:admissible1} that
there 
exists an {\em admissible} scaling $(\tilde{K}, \tilde{p}): \mathbb{N}_0 \to \mathbb{N}_0 \times [0,1]$
that satisfies (\ref{eq:ComparingAlpha+A}),
$\lim_{n \to \infty} (n-2\tilde{K}_n) = \infty$, $\limsup_{n \to \infty} \tilde{p}_n < 1$, 
and still 
the deviation function $\tilde{\gamma} : \mathbb{N}_0 \rightarrow
\mathbb{R}$ defined through
(\ref{eq:gamma_n_for_admissible})
 satisfying $\lim_{n \rightarrow \infty} \tilde \gamma_n = \infty$.
 Then, we get
 \[
 \lim_{n \rightarrow \infty } \bP{ \textrm{Min~node~degree~of~} 
\mathbb{H}\cap\mathbb{G}(n; \tilde{K}_n, \tilde{p_n})~\textrm{
is~no~less~than~} k} = 1
 \] 
from Proposition \ref{prop:nodedegree}, and the desired result
 \[
 \lim_{n \rightarrow \infty } \bP{ \textrm{Min~node~degree~of~} 
\mathbb{H}\cap\mathbb{G}(n; {K}_n, {p_n})~\textrm{
is~no~less~than~} k} = 1
 \] 
follows from (\ref{eq:coupling_OY_2}) since (\ref{eq:ComparingAlpha+A}) holds.

In a similar manner, pick any scaling $(K, p): \mathbb{N}_0 \rightarrow \mathbb{N}_0 \times 
  [0,1]$ such that 
$\lim_{n \to \infty} (n-2K_n) = \infty$, $\limsup_{n \to \infty} p_n < 1$. If the deviation sequence
$\gamma_n$ defined through (\ref{eq:scalinglaw}) satisfies $\lim_{n \to \infty} \gamma_n = - \infty$,
then we know from Proposition \ref{prop:admissible2} that
there 
exists an {\em admissible} scaling $(\tilde{K}, \tilde{p}): \mathbb{N}_0 \to \mathbb{N}_0 \times [0,1]$
that satisfies (\ref{eq:ComparingAlpha+A_2}),
$\lim_{n \to \infty} (n-2\tilde{K}_n) = \infty$, $\limsup_{n \to \infty} \tilde{p}_n < 1$, 
and still 
the deviation function $\tilde{\gamma} : \mathbb{N}_0 \rightarrow
\mathbb{R}$ defined through
(\ref{eq:gamma_n_for_admissible})
 satisfying $\lim_{n \rightarrow \infty} \tilde \gamma_n = - \infty$.
 Then, we get
 \[
 \lim_{n \rightarrow \infty } \bP{ \textrm{Min~node~degree~of~} 
\mathbb{H}\cap\mathbb{G}(n; \tilde{K}_n, \tilde{p_n})~\textrm{
is~no~less~than~} k} = 0
 \] 
from Proposition \ref{prop:nodedegree}, and the desired result
 \[
 \lim_{n \rightarrow \infty } \bP{ \textrm{Min~node~degree~of~} 
\mathbb{H}\cap\mathbb{G}(n; {K}_n, {p_n})~\textrm{
is~no~less~than~} k} = 0
 \] 
follows from (\ref{eq:coupling_OY_1}) since (\ref{eq:ComparingAlpha+A_2}) holds.

The rest of the paper will be devoted to establishing the next result; i.e.,  
Theorem \ref{thm:nodedegree} under admissible scalings.

\begin{proposition}
{\sl Consider an admissible scaling $(K, p): \mathbb{N}_0 \rightarrow \mathbb{N}_0 \times 
  [0,1]$ such that
$\lim_{n \to \infty} (n-2K_n) = \infty$ and $\limsup_{n \to \infty} p_n < 1$.
With the sequence $\gamma: \mathbb{N}_0 \rightarrow \mathbb{R}$
defined through (\ref{eq:scalinglaw}),
we have
\begin{eqnarray}
\lim_{n \rightarrow \infty } \bP{ \textrm{Min~node~degree~of~} \atop
\mathbb{H}\cap\mathbb{G}(n;\theta_n)~\textrm{
is~no~less~than~} k} = \left \{
\begin{array}{ll}
0 & \mbox{\textrm{if}~ $\lim\limits_{n \to \infty}\gamma_n = -\infty$} \\
  &                      \\
1 & \mbox{\textrm{if} ~$\lim\limits_{n \to \infty}\gamma_n = +\infty$}. 
\end{array}
\right . \label{eq:OneLaw+NodeIsolation_prop}
\end{eqnarray}
} \label{prop:nodedegree}
\end{proposition}

\subsection{Useful consequences of the scaling condition (\ref{eq:scalinglaw})}
\label{subsec:useful}

We collect in this section some useful consequences of the scaling condition (\ref{eq:scalinglaw})
that follow under the assumptions of admissibility and $\limsup_{n \to \infty} p_n < 1$.

\begin{lemma}
{\sl Consider an admissible scaling $(K, p): \mathbb{N}_0 \rightarrow \mathbb{N}_0 \times 
 [0,1]$ such that
$\limsup_{n \to \infty} p_n < 1$. Namely, 
the sequence $\gamma: \mathbb{N}_0 \rightarrow \mathbb{R}$
defined through (\ref{eq:scalinglaw}) satisfies $\gamma_n = \pm o(\log n)$.
Then, we have
\begin{equation}
p_n K_n = \Theta (\log n)
\label{eq:useful_p_n_K_n}
\end{equation}
and thus
\begin{equation}
K_n = \Omega(\log n).
\label{eq:useful_K_n}
\end{equation}
}
\label{lem:useful_bounds}
\end{lemma}
\myproof Under the admissibility condition $\gamma_n = \pm o(\log n)$, it is clear that 
 (\ref{eq:scalinglaw}) implies
 \begin{equation}
 p_nK_n\left(1-\frac{\log(1-p_n)}{p_n}-\frac{K_n}{n-1}\right)= \Theta(\log n). 
 \label{eq:to_useful_lemma_1}
 \end{equation}
 Next, observe that if $\limsup_{n \to \infty} p_n < 1$, then
 \[
1 \leq - \frac{\log(1-p_n)}{p_n} \leq M
 \]
 for some finite scalar $M$. Thus, it is immediate that
 \begin{equation}
1 \leq \left(1-\frac{\log(1-p_n)}{p_n}-\frac{K_n}{n-1}\right) \leq M+1
\label{eq:to_useful_lemma_2}
 \end{equation}
since $-1 \leq - \frac{K_n}{n-1} \leq 0$. The desired result (\ref{eq:useful_p_n_K_n}) is now immediate from 
 (\ref{eq:to_useful_lemma_1}) and (\ref{eq:to_useful_lemma_2}). Since $p_n \leq 1$ for all $n$,
(\ref{eq:useful_K_n}) follows clearly from (\ref{eq:useful_p_n_K_n}).
\myendpf

\section{A proof of Theorem \ref{thm:nodedegree}}
\label{sec:ProofTheoremNodeIsolation}

As the discussion in Section \ref{subsec:confining} shows, the proof of Theorem \ref{thm:nodedegree} 
will be completed if we establish Proposition \ref{prop:nodedegree}.
In this section, we outline the proof of Proposition \ref{prop:nodedegree} and then complete
the proof in several subsequent sections. 
The main idea behind the proof is to apply the method of
first and second moments \cite[p.  55]{JansonLuczakRucinski}
to the variable $X_{\ell} (n; \theta)$ that counts the number of nodes in $\mathbb{H}\cap\mathbb{G}(n;\theta)$ 
with degree $\ell$, for each
$\ell = 0,1 \ldots, n-1$. Namely, with $d_i$ denoting the degree of node $i$, i.e., 
\[
d_i = \sum_{j \in \{1,\ldots,n\} / \{i\}} \1{i \sim j},
\]
and $D_{i,\ell}$ standing for the event that node $i$ has degree $\ell$ (i.e., $D_{i,\ell}:=[d_i = \ell]$),
we set 
\begin{equation}
X_{\ell} (n; \theta) = \sum_{i=1}^{n} \1{D_{i, \ell}}.
\label{eq:X_ell}
\end{equation}
Note that the dependence
of the event $D_{i,\ell}$ (and, of the variable $d_i$) to the parameters $n$ and $\theta$ is suppressed here for notational convenience.
The graph $\mathbb{H}\cap\mathbb{G}(n;\theta)$ will have minimum node degree no less than $k$ if 
$X_{\ell} (n; \theta) = 0$ for each $\ell =0, 1, \ldots, k-1$. Similarly, the minimum node degree will be less than 
$k$ if for at least one of $\ell =0, 1, \ldots, k-1$, we have $X_{\ell} (n; \theta) > 0$.

Let $\delta(n; \theta)$ denote the minimum node degree in $\mathbb{H}\cap\mathbb{G}(n;\theta)$. 
The method of first and second moments \cite[p.  55]{JansonLuczakRucinski} will be used here in the forms 
\begin{equation}
\bP{X_{\ell} > 0} \leq \bE{ X_{\ell}}
\label{eq:FirstMoment}
\end{equation}
and
\begin{equation}
\frac{\bE{X_{\ell}}^2}{\bE{X_{\ell}^2}} \leq\bP{X_{\ell} > 0},
\label{eq:SecondMoment}
\end{equation}
respectively, which are valid for any positive-valued random variable $X_{\ell}$.

%
It is clear that collection of random variables $\1{D_{1, \ell}}, \ldots, \1{D_{n, \ell}}$ are exchangeable and thus we have
\begin{equation}
\bE{ X_{\ell} (n; \theta)} = n \bP{ D_{x,\ell} }
\label{eq:FirstMomentExpression}
\end{equation}
and
\begin{eqnarray}\label{eq:SecondMomentExpression}
\bE{ \left(X_{\ell} (n; \theta)\right)^2 } = n \bP{ D_{x,\ell} }
 +  n(n-1) \bP{ D_{x, \ell} \cap
D_{y,\ell} } \nonumber
\end{eqnarray}
by the binary nature of the rvs involved; here $x$ and $y$ are used to denote
generic node ids. It then follows that
\begin{eqnarray}
\frac{ \bE{ \left(X_{\ell}(n;\theta)\right)^2 }}{ \bE{ X_{\ell} (n;\theta) }^2 } 
= \frac{1}{ n\bP{ D_{x,\ell} } }
+ \frac{n-1}{n} \cdot \frac{\bP{ D_{x,\ell} \cap D_{y,\ell}  }}
     {\left (  \bP{ D_{x,\ell} } \right )^2 }.
\label{eq:SecondMomentRatio}
\end{eqnarray}

From (\ref{eq:FirstMoment}) and (\ref{eq:FirstMomentExpression})
it is plain that the one-law 
$\lim_{n \to \infty} \bP{\delta(n; \theta_n) \geq k}  =1 $
will be established if we show that
\begin{equation}
\lim_{n \to \infty } n \bP{ D_{x,\ell}} = 0 , \quad \ell = 0,1, \ldots, k-1  
\label{eq:OneLaw+NodeIsolation+convergence}
\end{equation}
From (\ref{eq:SecondMoment}) and
(\ref{eq:SecondMomentExpression}) we see that the zero-law 
$\lim_{n \to \infty} \bP{\delta(n; \theta_n) \geq k}  =0 $ holds if
\begin{equation}
\lim_{n \to \infty} n \bP{ D_{x,k-1} }= \infty
\label{eq:OneLaw+NodeIsolation+convergence2}
\end{equation}
and
\begin{equation}
\limsup_{n \to \infty}  \frac{\bP{ D_{x,k-1} \cap D_{y,k-1}  }}
     {\left (  \bP{ D_{x,k-1} } \right )^2 } \leq 1. \label{eq:ZeroLaw+NodeIsolation+convergence}
\end{equation}

The proof of Proposition \ref{prop:nodedegree} passes through
the next two technical propositions which establish
(\ref{eq:OneLaw+NodeIsolation+convergence}),
(\ref{eq:OneLaw+NodeIsolation+convergence2}) and
(\ref{eq:ZeroLaw+NodeIsolation+convergence}) under the appropriate
conditions on the scaling $\theta: \mathbb{N}_0 \rightarrow
\mathbb{N}_0 \times (0,1)$.

\begin{proposition}
{\sl Consider an admissible scaling $(K, p): \mathbb{N}_0 \rightarrow \mathbb{N}_0 \times 
  [0,1]$ such that
$\lim_{n \to \infty} (n-K_n) = \infty$ and $\limsup_{n \to \infty} p_n < 1$.
Define the sequence $\gamma_{\ell}: \mathbb{N}_0 \rightarrow \mathbb{R}$
 through 
 \begin{equation}
\label{eq:scalinglaw_general}
p_nK_n\left(1-\frac{\log(1-p_n)}{p_n}-\frac{K_n}{n-1}\right)= \log n + \ell \log \log n + \gamma_{\ell,n},
\end{equation}
for each $\ell =0,1, \ldots$, and for each $n=1,2, \ldots$.
Then, we have
\begin{eqnarray}
\lim_{n \rightarrow \infty } n \bP{ D_{x,\ell} }= \left \{
\begin{array}{ll}
0 & \mbox{\textrm{if}~ $\lim\limits_{n \to \infty}\gamma_{\ell,n} = +\infty$} \\
  &                      \\
\infty & \mbox{\textrm{if} ~$\lim\limits_{n \to \infty}\gamma_{\ell,n} = -\infty$}. 
\end{array}
\right . \label{eq:key_result_1}
\end{eqnarray}
} \label{prop:key_result_1}
\end{proposition}
A proof Proposition \ref{prop:key_result_1} is given in Section \ref{sec:proof_prop_1}.

\begin{proposition}
{\sl Consider an admissible scaling $(K, p): \mathbb{N}_0 \rightarrow \mathbb{N}_0 \times 
  [0,1]$ such that
$\lim_{n \to \infty} (n-2K_n) = \infty$ and $\limsup_{n \to \infty} p_n < 1$.
Then, we have
\begin{eqnarray}
 \bP{ D_{x,\ell} \cap D_{y,\ell}} \leq (1+o(1)) \left(\bP{ D_{x,\ell}}\right)^2 \label{eq:key_result_2}
\end{eqnarray}
for each $\ell=0,1, \ldots$.
} \label{prop:key_result_2}
\end{proposition}
A proof Proposition \ref{prop:key_result_2} can be found in Section \ref{sec:proof_prop_2}.

The proof of Proposition \ref{prop:nodedegree} can now be completed. 
Pick an admissible scaling $(K, p): \mathbb{N}_0 \rightarrow \mathbb{N}_0 \times 
  [0,1]$ such that
$\lim_{n \to \infty} (n-2K_n) = \infty$ and $\limsup_{n \to \infty} p_n < 1$.
Assume that the sequence $\gamma: \mathbb{N}_0 \rightarrow \mathbb{R}$
defined through (\ref{eq:scalinglaw}) satisfies $\lim_{n \to \infty }\gamma_n = + \infty$.
Comparing (\ref{eq:scalinglaw}) with (\ref{eq:scalinglaw_general}), this ensures that 
\[
\lim_{n \to \infty }\gamma_{\ell,n} = + \infty, \qquad \ell = 0,1,\ldots, k-1
\]
as we note that $\gamma_{\ell,n}$ is monotone decreasing in $\ell$ and that
$\lim_{n \to \infty }\gamma_n = + \infty$ is equivalent to $\lim_{n \to \infty }\gamma_{k-1,n} = + \infty$.
 It is clear that 
(\ref{eq:OneLaw+NodeIsolation+convergence}) follows by using
Proposition \ref{prop:key_result_1} for each $\ell = 0,1, \ldots, k-1$ and the one-law 
$\lim_{n \to \infty} \bP{\delta(n; \theta_n) \geq k}  =1 $ is immediate from
(\ref{eq:FirstMoment}) and (\ref{eq:FirstMomentExpression}).

Next, assume that the sequence $\gamma: \mathbb{N}_0 \rightarrow \mathbb{R}$
defined through (\ref{eq:scalinglaw}) satisfies $\lim_{n \to \infty }\gamma_n = - \infty$.
This is equivalent to $\lim_{n \to \infty }\gamma_{k-1,n} = - \infty$, and we get
(\ref{eq:OneLaw+NodeIsolation+convergence2}) from Proposition \ref{prop:key_result_1}
with $\ell = k-1$. Also, (\ref{eq:ZeroLaw+NodeIsolation+convergence}) follows from 
Proposition \ref{prop:key_result_2}, and the zero-law $\lim_{n \to \infty} \bP{\delta(n; \theta_n) \geq k}  =0 $
is now established via (\ref{eq:SecondMoment}) and
(\ref{eq:SecondMomentExpression}).
\myendpf

\section{A proof of Proposition \ref{prop:key_result_1}}
\label{sec:proof_prop_1}

\subsection{Outline}
For simplicity, we first consider the case when the parameters $K$ and $P$ are fixed; i.e., not scaled with $n$. 
The degree $d_x$ of an arbitrary node $x$ in $\mathbb{H \cap G}(n;\theta)$ can be written as the sum of two independent variables:
\[
d_x =  \sum_{j:~  j \in \Gamma_{n,x}(K)} \1{x \sim j} + \sum_{j:~ j \not \in \Gamma_{n,x}(K) ~ \textrm{and}~x \in \Gamma_{n,j}(K)} \1{x \sim j}
\]
where $\Gamma_{n,x}(K)$ and $\Gamma_{n,j}(K)$ are as defined previously in (\ref{eq:defn_of_Gamma}); i.e.,
they represent the set of $K$ nodes that are selected uniformly at random by nodes $x$ and $j$, respectively, in
the pairing mechanism of the random pairwise scheme. Notice that $x$ will not have an edge with $j$ in 
$\mathbb{H \cap G}(n;\theta)$ unless at least one of the events  $j \in \Gamma_{n,x}(K)$ and  $x \in \Gamma_{n,j}(K)$ 
takes place. Also, if either $j \in \Gamma_{n,x}(K)$ or $x \in \Gamma_{n,j}(K)$, the edge $x \sim j$ will exist in $\mathbb{H \cap G}(n;\theta)$ if and only if the communication channel between them is on; i.e., they are $B-$adjacent. Noting that
communication channels are {\em on} with probability $p$ independently from each other, we have 
 \begin{equation}
	d_x=_{\sl{st}} \textrm{Bin}\left(K ,p\right) +\textrm{Bin}\left(n-K-1,\frac{p K}{n-1}\right) 
\label{eq:probdist} 
\end{equation}
where $\textrm{Bin}(n,p)$ defines a standard {\em Binomial} distribution with $n$ trials 
and success probability $p$. This follows easily from the facts that
\[
|j:~  j \in \Gamma_{n,x}(K)| = K
\]
and
\[
|{j:~  j \not \in \Gamma_{n,x}(K) ~ \textrm{and}~x \in \Gamma_{n,j}(K)}| =_{\sl st} \textrm{Bin}\left(n-K-1, \frac{K}{n-1}\right),
\]
where the last relation follows from 
\begin{equation}
\bP{x \in \Gamma_{n,j}(K)} = {{{n-2}\choose{K-1}}\over{{n-1}\choose{K}}} = \frac{K}{n-1}, \qquad j \in \{1,\ldots,n \},~ j \neq x.
\label{eq:prob_x_in_y}
\end{equation}

The following result is the key in establishing Proposition \ref{prop:key_result_1} and will follow directly from the expression (\ref{eq:probdist}).
\begin{proposition}
{\sl Consider an admissible scaling $(K, p): \mathbb{N}_0 \rightarrow \mathbb{N}_0 \times 
  [0,1]$ such that
$\lim_{n \to \infty} (n-K_n) = \infty$ and $\limsup_{n \to \infty} p_n < 1$. For each $\ell=0,1,\ldots$, we have
\begin{eqnarray}\label{eq:probdist4}
	\bP{D_{x,\ell}} = (1+o(1)) \cdot \frac{(p_nK_n)^{\ell}}{\ell!}\cdot (1-p_n)^{K_n} \cdot \left(1-\frac{p_nK_n}{n-1}\right)^{n-K_n-1}
 \cdot \left(1-\frac{K_n}{n-1}+\frac{1}{1-p_n} \right)^{\ell}
\end{eqnarray}
}\label{prop:for_key_1}
\end{proposition}
The proof of Proposition \ref{prop:for_key_1} is given in Section \ref{subsec:proof_for_key_1}. 

We are now in a position to finish the proof of Proposition \ref{prop:key_result_1}. Consider an admissible scaling
 $(K, p): \mathbb{N}_0 \rightarrow \mathbb{N}_0 \times 
  [0,1]$ such that
$\lim_{n \to \infty} (n-K_n) = \infty$ and $\limsup_{n \to \infty} p_n < 1$. We consider each term in (\ref{eq:probdist4})
in turn. 
Since $K_n \leq n-1$ and $\limsup_{n \to \infty} p_n < 1$, we have
\begin{equation}
\left(1-\frac{K_n}{n-1}+\frac{1}{1-p_n} \right)^{\ell} = \Theta(1).
\label{eq:simplify_1_key_1}
\end{equation}
Also, in view of (\ref{eq:useful_p_n_K_n}), we have
\begin{equation}
(p_nK_n)^{\ell} = \Theta\left((\log n)^{\ell}\right), \qquad \ell=0,1, \ldots
\label{eq:simplify_2_key_1}
\end{equation}
Next, we make use of the following decomposition,
\begin{eqnarray}
	\log(1-x) = -x - \Psi(x), \quad \quad 0\leq x < 1
\label{eq:decomp} 
\end{eqnarray}
with
\begin{eqnarray}
	\Psi(x) = \int_0^x \frac{t}{1-t}dt
\end{eqnarray}
L'HospitalÕs rule yields $\lim_{x \downarrow 0} \frac{\Psi(x)}{x^2}=\frac{1}{2}$. Applying this decomposition, we get
\[
\left(1-\frac{p_nK_n}{n-1}\right)^{n-K_n-1} = \exp \left\{ - \frac{p_n K_n}{n-1}(n-K_n-1) - (n-K_n-1)\Psi\left(\frac{p_n K_n}{n-1}\right)\right\}
\]
Since $p_n K_n = \Theta(\log n)$ in view of Lemma \ref{lem:useful_bounds}, we have $\lim_{n \to \infty} \frac{p_n K_n}{n-1}=0$ and
$
\lim_{n \to \infty }\left(\frac{\Psi\left(\frac{p_n K_n} {n-1}\right)}{\frac{p_n^2 K_n^2}{(n-1)^2}}\right) =\frac{1}{2}.
$
This gives
\[
(n-K_n-1)\Psi\left(\frac{p_n K_n}{n-1}\right) = (n-K_n-1) \frac{p_n^2 K_n^2}{(n-1)^2} \left(\frac{\Psi\left(\frac{p_n K_n} {n-1}\right)}{\frac{p_n^2 K_n^2}{(n-1)^2}}\right) = O \left( \frac{(\log n)^2}{n-1}\right) = o(1).
\]
Thus, we conclude that
\begin{equation}
\left(1-\frac{p_nK_n}{n-1}\right)^{n-K_n-1} = \exp \left\{ -p_n K_n \left(1- \frac{K_n}{n-1}\right) + o(1)\right\}
 \label{eq:simplify_3_key_1}
\end{equation}

Finally, we report (\ref{eq:simplify_1_key_1}), (\ref{eq:simplify_2_key_1}), and (\ref{eq:simplify_3_key_1}) into 
(\ref{eq:probdist4}) to get 
\begin{equation}\nonumber
n \bP{D_{x,\ell}} = \Theta \left( \exp \left\{\log n + \ell \log \log n - p_n K_n \left(1- \frac{\log(1-p_n)}{p_n}-\frac{K_n}{n-1} \right) \right\}\right)
\end{equation}
We now invoke the scaling condition (\ref{eq:scalinglaw_general}) as in the statement of Proposition \ref{prop:key_result_1} to get
\[
n \bP{D_{x,\ell}} = \Theta \left( e^{-\gamma_{\ell,n}} \right) =e^{-\gamma_{\ell,n} + \Theta(1)}
\]
The desired result (\ref{eq:key_result_1}) is now immediate. \myendpf

\subsection{A proof of Proposition \ref{prop:for_key_1}}
\label{subsec:proof_for_key_1}
Recall that $\bP{D_{x,\ell}} = \bP{d_x = \ell}$. 
Given that the binomial distributions are independent in (\ref{eq:probdist}), we get 
\begin{eqnarray}\nonumber
	\bP{D_{x,\ell}} &=& \mathlarger{\sum}\limits_{i=0}^{\ell }{n-K-1 \choose i} \left(\frac{p K}{n-1}\right)^i \left(1-\frac{p K}{n-1}\right)^{n-K-1-i}
 \cdot {K \choose \ell-i}p^{\ell-i} (1-p)^{K-\ell+i}
 \\ \label{eq:probdist2} 
  &=& p^{\ell} (1-p)^K\left(1-\frac{p K}{n-1}\right)^{n-K-1} \mathlarger{\sum}\limits_{i=0}^{\ell}{n-K-1 \choose i} \left(\frac{K}{n-1}\right)^i \cdot{K \choose \ell-i}(1-p)^{i-\ell}
\nonumber \\
& &~ \cdot  \left(1-\frac{p K}{n-1}\right)^{-i} 
\end{eqnarray}
by an easy conditioning argument. 

Now, pick an admissible scaling $(K, p): \mathbb{N}_0 \rightarrow \mathbb{N}_0 \times 
  [0,1]$ such that
$\lim_{n \to \infty} (n-K_n) = \infty$ and $\limsup_{n \to \infty} p_n < 1$. We will invoke this scaling
into (\ref{eq:probdist2}). First, we introduce a simple asymptotic equivalency that will prove useful throughout:
For any sequence $a_n$ such that $\lim_{n \to \infty} a_n = \infty$ and any fixed scalar $i = 0,1, \ldots$, we have
\begin{equation}\label{eq:clm}
	{a_n \choose i}=\frac{a_n!}{(a_n-i)!~ i!} = \frac{a_n ^ i}{i!} \cdot \prod_{j=1}^{i-1}\left(1-\frac{j}{a_n}\right) =\frac{a_n^i}{i!}(1 + o(1)).
\end{equation}
Under the enforced assumptions, we have $K_n = \Omega(n)$ from Lemma \ref{lem:useful_bounds} so that $\lim_{n \to \infty} K_n = \infty$. Also, by assumption we have $\lim_{n \to \infty} (n-K_n) = \infty$. Thus, we can use (\ref{eq:clm}) in both of the combinatorial terms appearing in (\ref{eq:probdist2}).
Finally, under the enforced assumptions, we have $p_n K_n = \Theta(\log n)$ from Lemma \ref{lem:useful_bounds}, so that
\begin{equation}
\left(1-\frac{p_n K_n}{n-1}\right)^{-i} = 1+ o(1), \qquad i=0,1,\ldots, \ell.
\label{eq:1_minus_pK_to_zero}
\end{equation}

Reporting (\ref{eq:clm}) and (\ref{eq:1_minus_pK_to_zero}) together with the admissible scaling under consideration into (\ref{eq:probdist2}), we get
\begin{eqnarray}\label{eq:ap0} 
	\lefteqn{\bP{D_{x,\ell}}} \\ 
	 &=& (1+o(1)) p_n^{\ell} (1-p_n)^K\left(1-\frac{p_nK_n}{n-1}\right)^{n-K_n-1} \mathlarger{\sum}\limits_{i=0}^{\ell} \frac{(n-K_n-1)^i}{i!} \left(\frac{K_n}{n-1}\right)^i \frac{K_n^{\ell -i}}{(\ell-i)!}(1-p_n)^{i-\ell}
\nonumber \\ 
&=& (1+o(1)) \frac{(p_nK_n)^{\ell}}{\ell !} (1-p_n)^K\left(1-\frac{p_nK_n}{n-1}\right)^{n-K_n-1} \mathlarger{\sum}\limits_{i=0}^{\ell} {\ell \choose i} \left(\frac{n-K_n-1}{n-1}\right)^i (1-p_n)^{i-\ell}
\nonumber \\
&=& (1+o(1)) \frac{(p_nK_n)^{\ell}}{\ell !} (1-p_n)^K\left(1-\frac{p_nK_n}{n-1}\right)^{n-K_n-1} \left(1-\frac{K_n}{n-1}+\frac{1}{1-p_n}\right)^{\ell}
\end{eqnarray}
upon using Binomial Theorem in the last step. This completes the proof of Proposition \ref{prop:for_key_1}. \myendpf

\section{A proof of Proposition \ref{prop:key_result_2}}
\label{sec:proof_prop_2}

We start by obtaining an expression for $\bP{D_{x,\ell}\cap D_{y,\ell}}$. Qualitatively, this is the probability of two distinct nodes having degree $\ell$ in $\mathbb{H \cap G}(n;\theta)$. This probability clearly depends on whether or not there is an edge between $x$ and $y$ in $\mathbb{H \cap G}(n;\theta)$, which is tightly related to whether or not $x \sim_{K} y$; i.e., whether there is an edge between $x$ and $y$ in he individual $K$-out graph $\mathbb{H}(n;K)$. 
To that end, we use $\Gamma_x$ instead of $\Gamma_{n,x}(K)$ for any node $x$ to suppress the notation, and find it useful to 
write 
\begin{eqnarray}
	\bP{D_{x,\ell}\cap D_{y,\ell}} &=& \bP{D_{x,\ell}~\cap~ D_{y,\ell} ~\bigg |~ x\notin \Gamma_y ~,~ y\notin \Gamma_x } \cdot \bP{x\notin \Gamma_y ~\cap~ y\notin \Gamma_x}
	\nonumber\\
	&&~ +  2\cdot \bP{D_{x,\ell} ~\cap~ D_{y,\ell} ~\bigg |~ x\in \Gamma_y ~,~ y\notin \Gamma_x } \cdot \bP{x\in \Gamma_y ~\cap~ y\notin \Gamma_x}
	\nonumber\\
&&~ +  \bP{D_{x,\ell} ~\cap~ D_{y,\ell} ~\bigg |~ x\in \Gamma_y ~,~ y\in \Gamma_x } \cdot \bP{x\in \Gamma_y ~\cap~ y\in \Gamma_x}
\label{eq:second4} 
\end{eqnarray}
upon noting that by symmetry 
\[
\bP{D_{x,\ell} ~\cap~ D_{y,\ell} ~\bigg |~ x\in \Gamma_y ~,~ y\notin \Gamma_x } =
\bP{D_{x,\ell} ~\cap~ D_{y,\ell} ~\bigg |~ x\notin \Gamma_y ~,~ y\in \Gamma_x }.
\] 

Now, pick an admissible scaling $(K, p): \mathbb{N}_0 \rightarrow \mathbb{N}_0 \times 
 [0,1]$ such that $\lim_{n \to \infty} (n-2K_n) = \infty$ and $\limsup_{n \to \infty} p_n < 1$.
Note that for any node pair $x$ and $y$, we have 
$K_n \leq |\Gamma_x \cup \Gamma_y| \leq 2K_n$. Thus, conditioning on the event
$|\Gamma_x \cup \Gamma_y| = 2K_n-m$, the terms in (\ref{eq:second4}) can be written as follows
\begin{eqnarray}
\bP{D_{x,\ell}~\cap~ D_{y,\ell} ~\big |~ x\notin \Gamma_y , y\notin \Gamma_x }
&=& \sum\limits_{m = 0}^{K_n} \bP{D_{x,\ell}~\cap~ D_{y,\ell} ~\bigg|~ x \notin \Gamma_y , y\notin \Gamma_x , |\Gamma_x \cup \Gamma_y| = 2K_n-m}
\nonumber\\
& &~ \cdot \bP{|\Gamma_x \cup \Gamma_y| = 2K_n-m ~\big|~ x\notin \Gamma_y ~,~ y\notin \Gamma_x}
\label{eq:secondp01} 
\end{eqnarray}
\begin{eqnarray}
	\bP{D_{x,\ell}~\cap~ D_{y,\ell} ~\big |~ x\in \Gamma_y , y\notin \Gamma_x }
&=& \sum\limits_{m = 0}^{K_n} \bP{D_{x,\ell}~\cap~ D_{y,\ell} ~\bigg|~ x \in \Gamma_y , y\notin \Gamma_x , |\Gamma_x \cup \Gamma_y| = 2K_n-m}
\nonumber\\
& &~ \cdot \bP{|\Gamma_x \cup \Gamma_y| = 2K_n-m ~\big|~ x\in \Gamma_y ~,~ y\notin \Gamma_x}
\label{eq:secondp02} 
\end{eqnarray}
\begin{eqnarray}
	\bP{D_{x,\ell}~\cap~ D_{y,\ell} ~\big |~ x\in \Gamma_y , y\in \Gamma_x }
&=& \sum\limits_{m = 0}^{K_n} \bP{D_{x,\ell}~\cap~ D_{y,\ell} ~\bigg|~ x \in \Gamma_y , y\in \Gamma_x , |\Gamma_x \cup \Gamma_y| = 2K_n-m}
\nonumber\\
& &~ \cdot \bP{|\Gamma_x \cup \Gamma_y| = 2K_n-m ~\big|~ x\in \Gamma_y ~,~ y\in \Gamma_x}
\label{eq:secondp03} 
\end{eqnarray}

Proof of Proposition \ref{prop:key_result_2} passes through finding appropriate upper bounds for each of the terms 
(\ref{eq:secondp01}), (\ref{eq:secondp02}), and (\ref{eq:secondp03}). These bounds are provided in the next three
results, which are subsequently established in Sections \ref{sec:second1}, \ref{sec:second2}, and \ref{sec:second3}. For ease of notation, we define
\begin{eqnarray*}
	{P_1(n,\theta_n; m,\ell)} &:=& \bP{D_{x,\ell}~\cap~ D_{y,\ell} ~\bigg|~ x \notin \Gamma_y ~,~ y\notin \Gamma_x ~,~ |\Gamma_x \cup \Gamma_y| = 2K_n-m}
  \\
	{P_2(n,\theta_n; m,\ell)} &:=& \bP{D_{x,\ell}~\cap~ D_{y,\ell} ~\bigg|~ x \in \Gamma_y ~,~ y\notin \Gamma_x ~,~ |\Gamma_x \cup \Gamma_y| = 2K_n-m} \\
	{P_3(n,\theta_n; m,\ell)} &:=& \bP{D_{x,\ell}~\cap~ D_{y,\ell} ~\bigg|~ x \in \Gamma_y ~,~ y\in \Gamma_x ~,~ |\Gamma_x \cup \Gamma_y| = 2K_n-m}\end{eqnarray*}

\begin{proposition}
{\sl Consider an admissible scaling $(K, p): \mathbb{N}_0 \rightarrow \mathbb{N}_0 \times 
  [0,1]$ such that
$\lim_{n \to \infty} (n-2K_n) = \infty$ and $\limsup_{n \to \infty} p_n < 1$.
Given $\ell=0,1,\ldots$, we have
\begin{eqnarray}\label{eq:secondprop}
	{P_1(n,\theta_n; m,\ell)} \leq (1+o(1)) \bP{D_{x,\ell}}^2
\end{eqnarray}
for each $m=0, 1, \ldots,K_n$.
}\label{prop:second1}
\end{proposition}
A proof of proposition \ref{prop:second1} is given in Section \ref{sec:second1}.
\begin{proposition}
{\sl Consider an admissible scaling $(K, p): \mathbb{N}_0 \rightarrow \mathbb{N}_0 \times 
  [0,1]$ such that
$\lim_{n \to \infty} (n-2K_n) = \infty$ and $\limsup_{n \to \infty} p_n < 1$.
Given $\ell=0,1,\ldots$, we have
\begin{eqnarray}\label{eq:secondd2}
	{P_2(n,\theta_n; m,\ell)} \leq (1+o(1)) \bP{D_{x,\ell}}^2
\end{eqnarray}
for each $m=0,1,\ldots,K_n$.
}\label{prop:second2}
\end{proposition}
A proof of proposition \ref{prop:second2} is given in Section \ref{sec:second2}.

\begin{proposition}
{\sl Consider an admissible scaling $(K, p): \mathbb{N}_0 \rightarrow \mathbb{N}_0 \times 
  [0,1]$ such that
$\lim_{n \to \infty} (n-2K_n) = \infty$ and $\limsup_{n \to \infty} p_n < 1$. Given $\ell=0,1,\ldots$, we have
\begin{eqnarray}\label{eq:secondd3}
	{P_3(n,\theta_n; m,\ell)} \leq (1+o(1)) (1-p_n)^{-1} \bP{D_{x,\ell}}^2
\end{eqnarray}
for each $m=0, 1,\ldots,K_n$.
}\label{prop:second3}
\end{proposition}
A proof of proposition \ref{prop:second3} is given in Section \ref{sec:second3}.

We can now complete the proof of Proposition \ref{prop:key_result_2}. Reporting 
(\ref{eq:secondprop}), (\ref{eq:secondd2}), and (\ref{eq:secondd3}) for each $m=0, 1,\ldots,K_n$
into (\ref{eq:secondp01}), (\ref{eq:secondp02}), 
and (\ref{eq:secondp03}), respectively, we get 
\begin{eqnarray}
\bP{D_{x,\ell}~\cap~ D_{y,\ell} ~\big |~ x\notin \Gamma_y ~,~ y\notin \Gamma_x }
&\leq& (1+o(1)) \bP{D_{x,\ell}}^2
\label{eq:sp1final} \\
\bP{D_{x,\ell}~\cap~ D_{y,\ell} ~\big |~ x\in \Gamma_y ~,~ y\notin \Gamma_x }
&\leq& (1+o(1)) \bP{D_{x,\ell}}^2
\label{eq:sp2final} \\ 
\bP{D_{x,\ell}~\cap~ D_{y,\ell} ~\big |~ x\in \Gamma_y ~,~ y \in \Gamma_x }
&\leq& (1+o(1)) (1-p_n)^{-1} \bP{D_{x,\ell}}^2
\label{eq:sp3final} 
\end{eqnarray}
Now we use (\ref{eq:sp1final}), (\ref{eq:sp2final}), (\ref{eq:sp3final}) to bound $\bP{D_{x,\ell}\cap D_{y,\ell}}$ via (\ref{eq:second4}). 
It is clear that the desired result (\ref{eq:key_result_2}) will follow if we show that
\begin{eqnarray}
\bP{x\notin \Gamma_y ~\cap~ y\notin \Gamma_x} + 2 \cdot \bP{x\in \Gamma_y ~\cap~ y\notin \Gamma_x}
+ (1-p_n)^{-1} \bP{x\in \Gamma_y ~\cap~ y\in \Gamma_x} = 1+o(1)
\label{eq:finals1} 
\end{eqnarray}
under the enforced assumptions. 
Recalling (\ref{eq:prob_x_in_y}) and independence of $\Gamma_x$ and $\Gamma_y$, we get by a direct computation
that
\begin{eqnarray}\nonumber
\lefteqn{\bP{x\notin \Gamma_y \cap y\notin \Gamma_x} + 2 \cdot \bP{x\in \Gamma_y \cap y\notin \Gamma_x}
+ (1-p_n)^{-1} \bP{x\in \Gamma_y \cap y\in \Gamma_x}} && \\ 
 &=&  \left(1-\frac{K_n}{n-1}\right)^2 +  2\frac{K_n}{n-1}\left(1-\frac{K_n}{n-1}\right)+ (1-p_n)^{-1} \left(\frac{K_n}{n-1}\right)^2
\nonumber\\
&=& 1+\frac{K_n^2}{(n-1)^2} \left(\frac{1}{1-p_n}-1\right)
\nonumber\\
&=& 1+\frac{K_n^2 ~ p_n}{(n-1)^2~(1-p_n)} 
\label{eq:finals3} 
\end{eqnarray}
Note that we have $K_n \leq n-1$ and $p_n K_n = \Theta(\log n)$ by the admissibility of the
scaling; just recall (\ref{eq:ScalingDefn}) and (\ref{eq:useful_p_n_K_n}). We also have
$\limsup_{n \to \infty} p_n < 1$ by assumption. Combining we get 
\[
\frac{K_n^2 ~ p_n}{(n-1)^2~(1-p_n)} \leq \frac{K_n ~ p_n}{(n-1)} \cdot \frac{1}{1-p_n} = o(1)
\]
and (\ref{eq:finals1}) follows from (\ref{eq:finals3}). The desired result (\ref{eq:key_result_2}) is established
and the proof of Proposition \ref{prop:key_result_2} is now complete.
\myendpf

\section{A proof of Proposition \ref{prop:second1}}
\label{sec:second1}

\begin{figure}[!t]
\centering
\includegraphics[totalheight=0.22\textheight,
width=.5\textwidth]{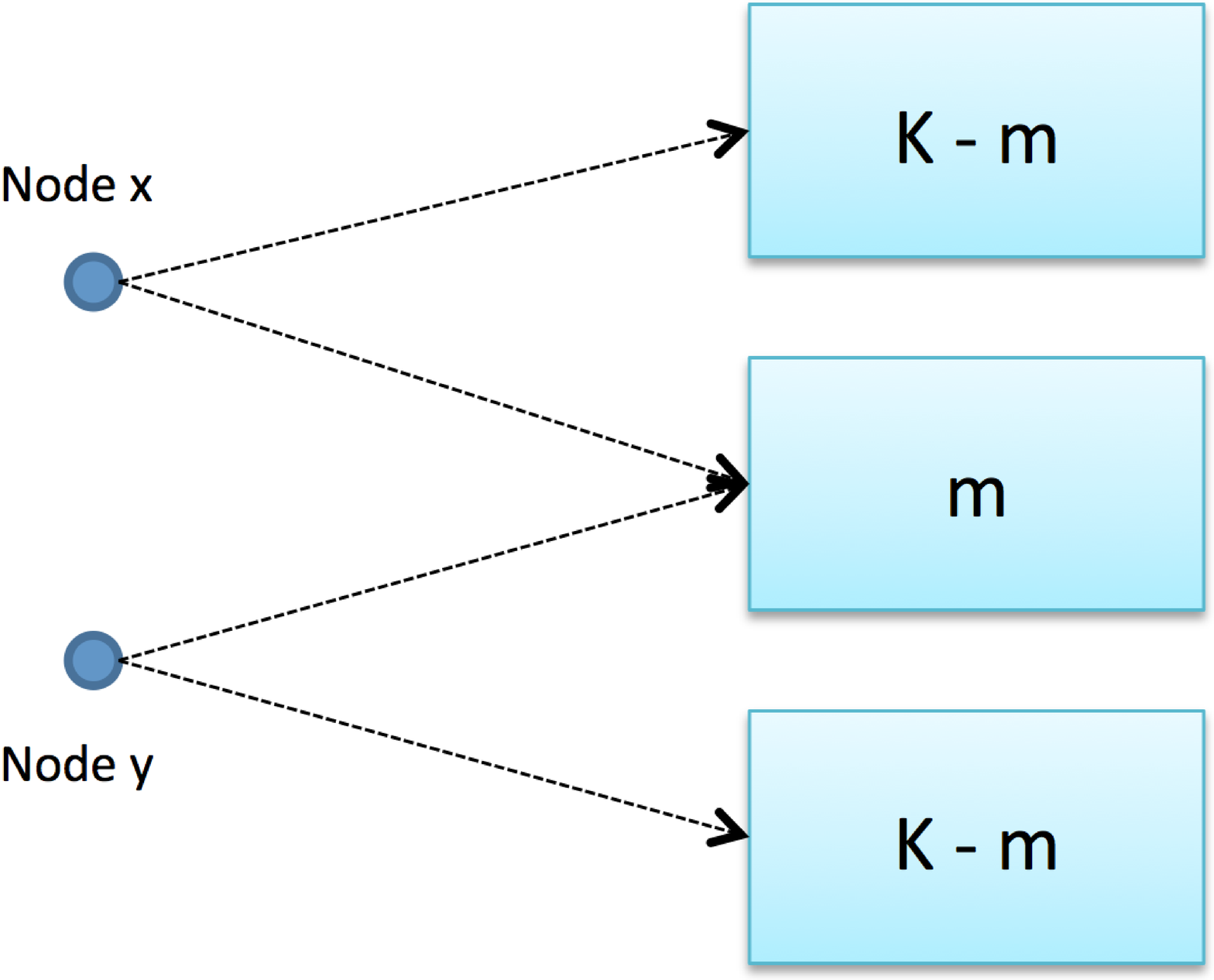} 
\caption{\sl Depicting the condition 
$(x \notin \Gamma_y ~,~ y\notin \Gamma_x ~,~ |\Gamma_x \cup \Gamma_y| = 2K_n-m)$ for the calculation
of $P_1(n,\theta_n; m, \ell)$. Dashed lines emanating from
a node $x$ stand for the set of nodes in $\Gamma_x$.}
\label{fig:p1}
\end{figure}

We will seek an exact expression for $P_1(n,\theta_n; m, \ell)$ first, and then apply judicious bounding arguments to get the desired result (\ref{eq:secondprop}). First, observe that under the condition $(x \notin \Gamma_y ~,~ y\notin \Gamma_x ~,~ |\Gamma_x \cup \Gamma_y| = 2K_n-m)$, nodes $x$ and $y$ do not have an edge in between (i.e., the event $(x \sim y)^{c}$ takes place) and all of their $\ell$ neighbors have to be among the $n-2$ nodes in $\mathcal{V}/\{x,y\}$. Furthermore, it is clear that
\begin{equation}
|\Gamma_x \cap \Gamma_y| = m, \qquad |\Gamma_x / \Gamma_y| = K_n-m, \qquad \textrm{and} \qquad |\Gamma_y / \Gamma_x| = K_n-m
\nonumber
\end{equation}
This situation is depicted in Figure \ref{fig:p1}. When calculating the probability $\bP{D_{x,\ell}~\cap~ D_{y,\ell}}$ under this condition, we first consider the possible neighbors of nodes $x$ and $y$ in the set $\Gamma_x \cup \Gamma_y$. Let $d_x (\Gamma_x \cup \Gamma_y)$ denote the number of neighbors of $x$ in $\mathbb{H \cap G}(n;\theta_n)$ restricted to node set $\Gamma_x \cup \Gamma_y$. More precisely, we set
\[
d_x (\Gamma_x \cup \Gamma_y) =  \sum_{z:~  z \in (\Gamma_x \cup \Gamma_y)} \1{x \sim z} 
\]
We define $d_y (\Gamma_x \cup \Gamma_y)$ similarly. Similar to (\ref{eq:probdist}), it is easy to check that
\[
d_x (\Gamma_x \cup \Gamma_y) = d_x (\Gamma_x) + d_x (\Gamma_y / \Gamma_x)
\]
with $d_x (\Gamma_x)$ and $d_x (\Gamma_y / \Gamma_x)$ independent, and 
\[
d_x (\Gamma_x) =_{\sl st} \textrm{Bin}(K_n,p_n) \qquad \textrm{and} \qquad 
d_x (\Gamma_y / \Gamma_x)=_{\sl st} \textrm{Bin}\left(K_n-m, \frac{p_n K_n}{n-1}\right)
\] 
Similar arguments
hold for the corresponding quantities for node $y$. In particular, we have
\[
d_x (\Gamma_x \cup \Gamma_y) =_{\sl st} d_y (\Gamma_x \cup \Gamma_y) =_{\sl st}\textrm{Bin}(K_n,p_n) + \textrm{Bin}\left(K_n-m, \frac{p_n K_n}{n-1}\right),
\]
and clearly $d_x (\Gamma_x \cup \Gamma_y)$ and $d_y (\Gamma_x \cup \Gamma_y)$
are independent. 

Next, we have to consider the possible neighbors of $x$ and $y$ among the $n-(2K_n-m+2)$ nodes in 
$\mathcal{V}/(\{x,y\}\cup \Gamma_x \cup \Gamma_y)$. This time, $d_x(\mathcal{V}/(\{x,y\}\cup \Gamma_x \cup \Gamma_y))$ and $d_y(\mathcal{V}/(\{x,y\}\cup \Gamma_x \cup \Gamma_y))$ are {\em not} independent from each other. In fact, they are {\em negatively associated} in the sense of Joag-Dev and Proschan \cite{JoagdevProschan}. The reader is referred to
\cite[Section IX]{YaganMakowskiPER} for a formal proof of this claim, but it 
is a consequence of the fact that the set $\Gamma_z$ for any node $z$ is a random sample ({\em without replacement}) of size $K$ from 
$\mathcal{V}/\{z\}$.
This fact will be exploited here in the following way. Note that
each node $z$ in $\mathcal{V}/(\{x,y\}\cup \Gamma_x \cup \Gamma_y)$ will satisfy one of the following {\em independently} 
from any other node in $\mathcal{V}/(\{x,y\}\cup \Gamma_x \cup \Gamma_y)$: 
\begin{itemize}
\item[i)] $(z \sim x) \cap (z \sim y)$, with probability $\bP{(z \sim x) \cap (z \sim y)} \leq (\bP{(z \sim x)})^2 $ 
\item[ii)] $(z \sim x)^{c} \cap (z \sim y)$ with probability $\bP{(z \sim x)^c \cap (z \sim y)} \leq \bP{(z \sim y)} $ 
\item[iii)] $(z \sim x) \cap (z \sim y)^c$, with probability $\bP{(z \sim x) \cap (z \sim y)^c} \leq \bP{(z \sim x)} $ 
\item[iv)] $(z \sim x)^c \cap (z \sim y)^c$ with probability $\bP{(z \sim x)^{c} \cap (z \sim y)^{c}} \leq (\bP{(z \sim x)^{c}})^2 $ 
\end{itemize}
The bound in item (i) follows from the negative association of the events $(z \sim x)$ and $(z \sim y)$, which also implies the negative association of $(z \sim x)^{c}$ and $(z \sim y)^{c}$ leading to the bound in item (iv). The bounds in items (ii) and
(iii) hold trivially.

Combining these arguments, we now get
\begin{eqnarray}
\lefteqn{P_1(n;\theta_n;m,\ell)} &&
\nonumber \\
 &=& \sum\limits_{i,j = 0}^{\ell}{K_n \choose i}{K_n \choose j}  p_n^{i+j}(1-p_n)^{2K_n-i-j} ~\sum\limits_{i_1,j_1 = 0}^{\ell-i,\ell-j}{K_n-m \choose i_1}{K_n-m \choose j_1} \cdot \left(\frac{p_nK_n}{n-1} \right)^{i_1+j_1}
\nonumber\\
& &~ \left(1-\frac{p_nK_n}{n-1} \right)^{2K_n-2m-i_1-j_1}~ \sum\limits_{u = 0}^{\ell-\max{(i+i_1,j+j_1)}}{n-2K_n+m-2 \choose u}
(\bP{(z \sim x) \cap (z \sim y)})^u
\nonumber\\
& &~ \cdot {n-2K_n+m-2-u \choose \ell-i-i_1-u} (\bP{(z \sim x) \cap (z \sim y)^c}) ^{\ell-u-i-i_1} 
\cdot {n-2K_n+m-2-\ell+i+i_1 \choose \ell-j-j_1-u}
\nonumber\\
& &~ \cdot(\bP{(z \sim x)^c \cap (z \sim y)}) ^{\ell-u-j-j_1} 
\left(\bP{(z \sim x)^c \cap (z \sim y)^c} \right)^{n-2K_n+m-2-2\ell+i+i_1+j+j_1+u}
\label{eq:secondp1final}
\\
&\leq& \sum\limits_{i,j = 0}^{\ell}{K_n \choose i}{K_n \choose j}  p_n^{i+j}(1-p_n)^{2K_n-i-j} ~ \sum\limits_{i_1,j_1 = 0}^{\ell-i,\ell-j}{K_n-m \choose i_1}{K_n-m \choose j_1} \cdot \left(\frac{p_nK_n}{n-1} \right)^{i_1+j_1}
\nonumber\\
& & \cdot \left(1-\frac{p_nK_n}{n-1} \right)^{2K_n-2m-i_1-j_1} ~ \sum\limits_{u = 0}^{\ell-\max{(i+i_1,j+j_1)}}{n-2K_n+m-2 \choose u} \left(\frac{p_nK_n}{n-1} \right)^{2u}
\nonumber\\
& & \cdot  {n-2K_n+m-2-u \choose \ell-i-i_1-u}\left(\frac{p_nK_n}{n-1} \right)^{2\ell-2u-i-i_1-j-j_1} \nonumber\\
& & \cdot {n-2K_n+m-2-\ell+i+i_1 \choose \ell-j-j_1-u}
 \left( 1-\frac{p_nK_n}{n-1} \right)^{2(n-2K_n+m-2-2\ell+i+i_1+j+j_1+u)}
\label{eq:p1p1} 
\end{eqnarray}
with $z$ denoting an arbitrary node in $\mathcal{V}/(\{x,y\}\cup \Gamma_x \cup \Gamma_y)$.
In (\ref{eq:secondp1final}), we used a series of conditioning arguments with the following notation: 
$d_x (\Gamma_x) = i$, $d_y (\Gamma_y) = j$, $d_x (\Gamma_y / \Gamma_x) = i_1$, $d_y (\Gamma_x / \Gamma_y) = j_1$, 
and $u$ denoting the number of nodes in $\mathcal{V}/(\{x,y\}\cup \Gamma_x \cup \Gamma_y)$ that are connected
to both $x$ and $y$; i.e., $u= |\{z \in \mathcal{V}/(\{x,y\}\cup \Gamma_x \cup \Gamma_y):~ (z\sim x)\cap(z \sim y)\}|$.
Also, (\ref{eq:p1p1}) follows easily from the bounds introduced in items (i)-(iv) above and from the fact that
\[
\bP{(z \sim x) ~|~ z \not \in \Gamma_x} = \bP{ (x \in \Gamma_z) ~\cap~ (B_{xz}(p_n)=1)} = \frac{K_n}{n-1} p_n.
\]

We now simplify this bound further. We first apply available cancelations and then use (\ref{eq:clm}) for ${K_n \choose i}$ and ${K_n \choose j}$ since $K_n =\Omega(\log n)$ under the enforced assumptions; just recall (\ref{eq:useful_K_n}). 
Finally, multiplying and dividing by $\ell!$, we get the following simplified version.
\begin{eqnarray}
\lefteqn{P_1(n,\theta_n;m,\ell)} &&
\\&\leq& (1+o(1))(p_nK_n)^{2\ell}(1-p_n)^{2K_n}\left(1-\frac{p_nK_n}{n-1}\right)^{2(n-K_n-1)}
\left(\frac{1}{\ell!}\right)^2
\nonumber\\
& &~ \cdot \sum\limits_{i,j = 0}^{\ell}\left(\frac{K_n^i K_n^j(\ell!)^2}{i!j!}\right)  p_n^{i+j}(1-p_n)^{-i-j}(p_nK_n)^{-i-j}\left(\frac{1}{n-1}  \right)^{2\ell -i-j} ~\sum\limits_{i_1,j_1 = 0}^{\ell-i,\ell-j}{K_n-m \choose i_1}
\nonumber\\
& &~ \cdot {K_n-m\choose j_1}  \cdot \sum\limits_{u = 0}^{\ell-\max{(i+i_1,j+j_1)}}{n-2K_n+m-2 \choose u} 
\left( 1-\frac{p_nK_n}{n-1} \right)^{-2-4\ell+2i+2j+2u+i_1+j_1}
 \nonumber\\
& &~ \cdot   {n-2K_n+m-2-u \choose \ell-i-i_1-u}{n-2K_n+m-2-\ell+i+i_1 \choose \ell-j-j_1-u}
\label{eq:p1p2} 
\end{eqnarray}
Note that $\ell,u,i,j,i_1,j_1$ are all bounded constants. Thus, in view of (\ref{eq:useful_p_n_K_n}), we clearly have
\begin{eqnarray}
	\left( 1-\frac{p_nK_n}{n-1} \right)^{-2-4\ell+2i+2j+2u+i_1+j_1} = 1 + o(1)
\label{eq:p1p3} 
\end{eqnarray}
Using this, and recalling (\ref{eq:probdist4}), we see that Proposition \ref{prop:second1} will be established 
if we show that
\begin{eqnarray}
\lefteqn{\sum\limits_{i,j = 0}^{\ell}\left(\frac{(\ell!)^2}{i!j!}\right)  \left(\frac{1}{n-1}\right)^{2\ell -i-j}(1-p_n)^{-i-j}\sum\limits_{i_1,j_1 = 0}^{\ell-i,\ell-j}{K_n-m \choose i_1} {K_n-m\choose j_1} } &&
\label{eq:for_p_1_to_show_1}\\
&&\cdot 
\hspace{-4mm}\sum\limits_{u = 0}^{\ell-\max{(i+i_1,j+j_1)}}{n-2K_n+m-2 \choose u}  {n-2K_n+m-2-u \choose \ell-i-i_1-u} {n-2K_n+m-2-\ell+i+i_1 \choose \ell-j-j_1-u}
\nonumber\\
&\leq&  (1+o(1)) \left(1-\frac{K_n}{n-1}+\frac{1}{1-p_n} \right)^{2 \ell}
\label{eq:p1p4} 
\end{eqnarray}
for each $m=0,1,\ldots,K_n$.

Under the enforced assumption that $\lim_{n \to \infty}(n-2K_n) = \infty$, we have for any pair of constants $c_1$, $c_2$ that
\begin{eqnarray}\nonumber
{n-2K_n+m \pm c_1 \choose c_2} \leq \frac{(n-2K_n+m \pm c_1)^{c_2}}{c_2!} &=& \frac{(n-2K_n+m)^{c_2}}{c_2!} \left(1 \pm \frac{c_1}{n-2K_n+m}\right)^{c_2} 
\nonumber \\
&=& (1+o(1)) \frac{(n-2K_n+m)^{c_2}}{c_2!} 
\label{eq:useful_bound_combinatorials}
\end{eqnarray}
In view of this, we get
\begin{eqnarray}
\lefteqn{\sum\limits_{u = 0}^{\ell-\max{(i+i_1,j+j_1)}}{n-2K_n+m-2 \choose u}  {n-2K_n+m-2-u \choose \ell-i-i_1-u} {n-2K_n+m-2-\ell+i+i_1 \choose \ell-j-j_1-u}} \hspace{5mm}&&
\nonumber \\
&\leq& (1+o(1)) \sum\limits_{u = 0}^{\ell-\max{(i+i_1,j+j_1)}} \frac{(n-2K_n+m)^{2\ell-i-i_1-j-j_1-u}}{u!(\ell-i-i_1-u)!(\ell-j-j_1-u)!}
~~~~~~~~~~~~~~~~~~~~\qquad\qquad
\nonumber \\
&=& (1+o(1)) (n-2K_n+m)^{2\ell-i-i_1-j-j_1} \sum\limits_{u = 0}^{\ell-\max{(i+i_1,j+j_1)}} \frac{(n-2K_n+m)^{-u}}{u!(\ell-i-i_1-u)!(\ell-j-j_1-u)!}
\nonumber \\
&=& (1+o(1)) (n-2K_n+m)^{2\ell-i-i_1-j-j_1} \cdot \frac{1}{(\ell-i-i_1)!(\ell-j-j_1)!}
\label{eq:p1p9} 
\end{eqnarray}
upon noting that
\begin{eqnarray}
\lefteqn{\sum\limits_{u = 0}^{\ell-\max{(i+i_1,j+j_1)}} \frac{(n-2K_n+m)^{-u}}{u!(\ell-i-i_1-u)!(\ell-j-j_1-u)!}} && 
\nonumber \\
&=& \frac{1}{(\ell-i-i_1)!(\ell-j-j_1)!} + \sum\limits_{u = 1}^{\ell-\max{(i+i_1,j+j_1)}} \frac{(n-2K_n+m)^{-u}}{u!(\ell-i-i_1-u)!(\ell-j-j_1-u)!}
\nonumber \\
&=& \frac{1}{(\ell-i-i_1)!(\ell-j-j_1)!} + o(1)
\nonumber \\
&=& \frac{1}{(\ell-i-i_1)!(\ell-j-j_1)!} (1+o(1))
\nonumber
\end{eqnarray}
in view of $\lim_{n \to \infty} (n-2K_n+m) = \infty$ for all $m=0,1,\ldots,K_n$.

We now report (\ref{eq:p1p9}) into (\ref{eq:for_p_1_to_show_1}) and note that
\begin{eqnarray}
\lefteqn{\sum\limits_{i_1,j_1 = 0}^{\ell-i,\ell-j}{K_n-m \choose i_1} {K_n-m\choose j_1}  (n-2K_n+m)^{2\ell-i-i_1-j-j_1} \cdot 
\frac{1}{(\ell-i-i_1)!(\ell-j-j_1)!}} &&
\nonumber \\
&\leq&  \sum\limits_{i_1,j_1 = 0}^{\ell-i,\ell-j} \frac{(K_n-m)^{i_1+j_1}}{i_1! ~ j_1!} \frac{(n-2K_n+m)^{2\ell-i-j-i_1-j_1}} {(\ell-i-i_1)!(\ell-j-j_1)!} 
\nonumber \\ 
&=&  \frac{1}{(\ell - i)! ~(\ell - j)!} \cdot \sum\limits_{i_1,j_1 = 0}^{\ell-i,\ell-j} {{\ell -i} \choose i_1} 
{{\ell -j} \choose j_1}
\left(K_n-m\right)^{i_1+j_1} \left(n-2K_n+m\right)^{2\ell-i-j-i_1-j_1}
\nonumber \\ 
&=&  \frac{1}{(\ell - i)! ~(\ell - j)!} \cdot \left(\sum\limits_{i_1= 0}^{\ell-i} {{\ell -i} \choose i_1} 
\left(K_n-m\right)^{i_1} \left(n-2K_n+m\right)^{\ell-i-i_1} \right)
\nonumber \\
& & ~ \cdot  \left(\sum\limits_{j_1 =0}^{\ell-j} {{\ell -j} \choose j_1} 
\left(K_n-m\right)^{j_1} \left(n-2K_n+m\right)^{\ell-j-j_1} \right)
\nonumber \\ 
&=&\frac{1}{(\ell - i)! ~(\ell - j)!}\cdot \left(n-K_n  \right)^{2\ell-i - j}
\label{eq:int_p1p11}
\end{eqnarray}
upon using Binomial Theorem in the last step.
Using (\ref{eq:p1p9}) together with (\ref{eq:int_p1p11}) in (\ref{eq:for_p_1_to_show_1}), we get
\begin{eqnarray}
\lefteqn{\sum\limits_{i,j = 0}^{\ell}\left(\frac{(\ell!)^2}{i!j!}\right)  \left(\frac{1}{n-1}\right)^{2\ell-i-j}(1-p_n)^{-i-j}\sum\limits_{i_1,j_1 = 0}^{\ell-i,\ell-j}{K_n-m \choose i_1} {K_n-m\choose j_1} } &&
\\
&&\cdot 
\hspace{-4mm}\sum\limits_{u = 0}^{\ell-\max{(i+i_1,j+j_1)}}{n-2K_n+m-2 \choose u}  {n-2K_n+m-2-u \choose \ell-i-i_1-u} {n-2K_n+m-2-\ell+i+i_1 \choose \ell-j-j_1-u}
\nonumber\\ \nonumber
&\leq&
(1+o(1)) \sum\limits_{i,j = 0}^{\ell}\left(\frac{(\ell!)^2}{i!j!}\right)  \left(\frac{1}{n-1}\right)^{2\ell -i-j}(1-p_n)^{-i-j}
\frac{1}{(\ell - i)! ~(\ell - j)!} \cdot \left(n-K_n \right)^{2\ell-i-j}
\nonumber \\
&=&(1+o(1)) \left(\sum\limits_{i = 0}^{\ell} {\ell \choose i} \left((1-p_n)^{-1}\right)^{i} \left(\frac{n-K_n}{n-1}\right)^{\ell-i}\right)^{2}
\nonumber \\
&=&(1+o(1)) \left(1- \frac{K_n-1}{n-1} + \frac{1}{1-p_n} \right)^{2\ell}
\nonumber \\
&=&(1+o(1)) \left(1- \frac{K_n}{n-1} + \frac{1}{1-p_n} + o(1) \right)^{2\ell}
\nonumber \\
&=&(1+o(1)) \left(1- \frac{K_n}{n-1} + \frac{1}{1-p_n}  \right)^{2\ell},
\nonumber
\end{eqnarray}
where in the last step we used the fact that
\begin{equation}
\left(1- \frac{K_n}{n-1} + \frac{1}{1-p_n}  \right) \geq 1
\label{eq:extra_term_lower_bound}
\end{equation}
in view of (\ref{eq:ScalingDefn}).
Thus, we get the desired result (\ref{eq:p1p4}) for each $m=0,1,\ldots,K_n$ and the proof
of Proposition \ref{prop:second1} is now completed.
\myendpf

\section{A proof of Proposition \ref{prop:second2}}
\label{sec:second2}

We first condition on the event $(x \sim_{B} y)$ and write $P_2(n,\theta_n; m, \ell)$ as
\begin{eqnarray}
	P_2(n,\theta_n; m, \ell) = p_n\cdot P_{21}(n,\theta_n; m, \ell) + (1-p_n) \cdot P_{22}(n,\theta_n; m, \ell)
\label{eq:p2opened}
\end{eqnarray}
with $P_{21}(n,\theta_n; m, \ell)$ and $P_{22}(n,\theta_n; m, \ell)$ defined through
\begin{eqnarray}
	P_{21}(n,\theta_n; m, \ell) &=& \bP{D_{x,\ell}\cap D_{y,\ell} ~\big |~ x\in \Gamma_y ~,~ y\notin \Gamma_x ~,~ |\Gamma_x \cup \Gamma_y| = 2K_n-m ~,~ B_{xy}(p_n)=1}
\label{eq:p21} \nonumber
	\\
		P_{22}(n,\theta_n; m, \ell) &=& \bP{D_{x,\ell}\cap D_{y,\ell} ~\big|~ x\in \Gamma_y ~,~ y\notin \Gamma_x ~,~ |\Gamma_x \cup \Gamma_y| = 2K_n-m ~,~ B_{xy}(p_n)=0}
\label{eq:p22} \nonumber
\end{eqnarray}
Recall that $B_{xy}(p_n)$ is defined in Section \ref{subsec:intersection_random_graphs} and controls whether
the wireless channel between nodes $x$ and $y$ is {\em on} ($B_{xy}(p_n)=1$), or {\em off} ($B_{xy}(p_n)=0$). 
Thus, (\ref{eq:p2opened}) follows upon noting that $B_{xy}(p_n)$ is independent from $\Gamma_x$ and $\Gamma_y$.

\begin{figure}[!t]
\centering
\includegraphics[totalheight=0.22\textheight,
width=.5\textwidth]{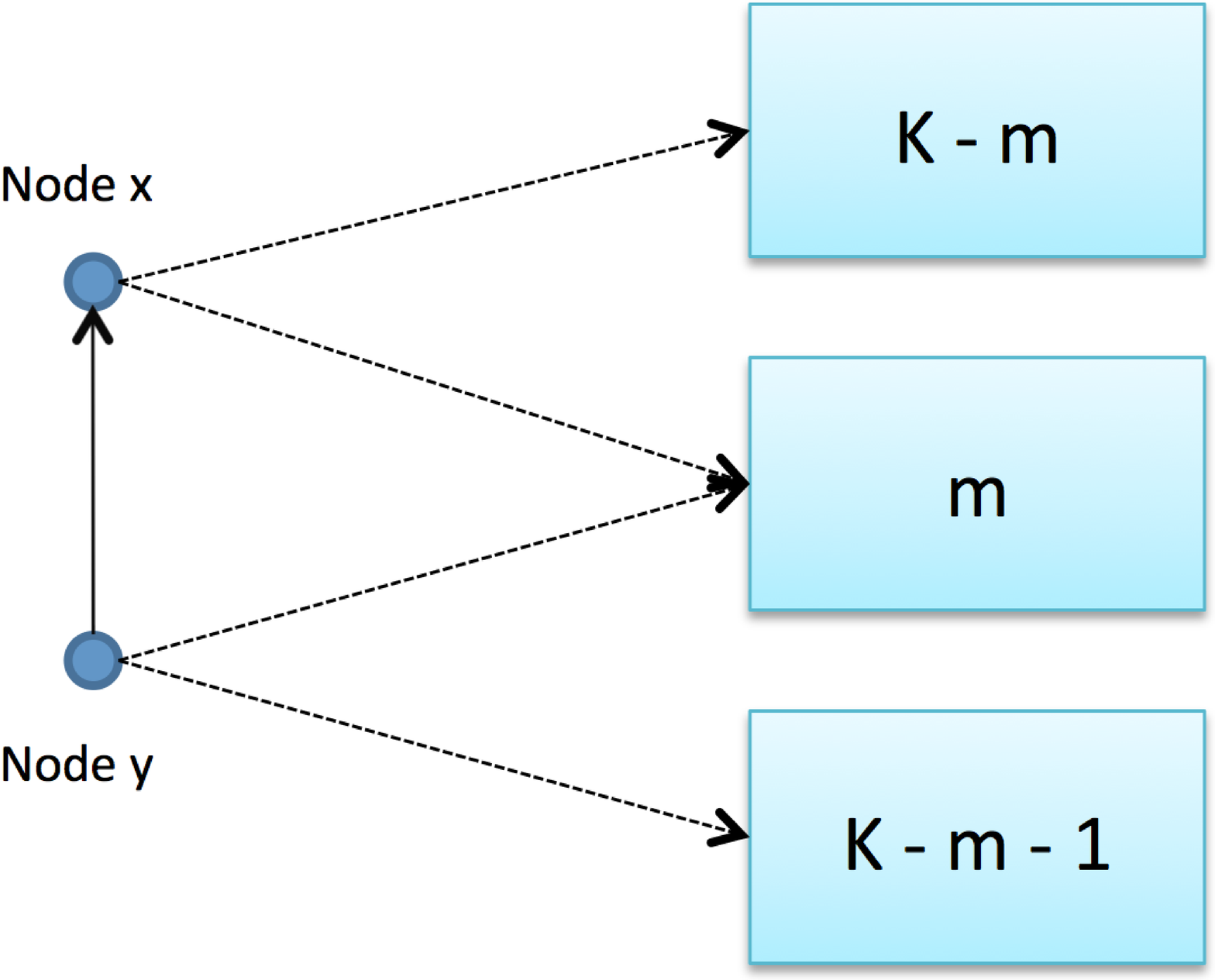} 
\caption{\sl Depicting the condition 
$(x \in \Gamma_y ~,~ y\notin \Gamma_x ~,~ |\Gamma_x \cup \Gamma_y| = 2K_n-m ~,~ B_{xy}=1)$ for the calculation
of $P_{21}(n,\theta_n; m, \ell)$. Dashed lines emanating from
a node $x$ stand for the set of nodes in $\Gamma_x$ and a dashed line between $x$ and $y$
is made bold if $B_{xy}=1$.}
\label{fig:p2}
\end{figure}

We will consider each of the terms $P_{21}(n,\theta_n; m, \ell)$ and $P_{22}(n,\theta_n; m, \ell)$ separately. 
We start by showing that
\begin{equation}
P_{21}(n,\theta_n; m, \ell) = o\left(\bP{D_{x,\ell}}^2 \right), \qquad \ell=1, 2, \ldots
\label{eq:p_21_zero}
\end{equation}
for each $m=0,1,\ldots, K_n$ under the enforced assumptions. First, note that under the condition $(x \in \Gamma_y ~,~ y\notin \Gamma_x ~,~ |\Gamma_x \cup \Gamma_y| = 2K_n-m~,~ B_{xy}(p_n)=1)$, nodes $x$ and $y$ do have an edge in between (in the intersection graph
$\mathbb{H}\cap\mathbb{G}(n;\theta_n)$), and they just need to have $\ell-1$ additional neighbors among the $n-2$ nodes in $\mathcal{V}/\{x,y\}$. Furthermore, it is clear that
\begin{equation}
|\Gamma_x \cap \Gamma_y| = m, \quad |\Gamma_x / (\Gamma_y \cup \{x,y\})| = K_n-m, \quad \textrm{and} \quad |\Gamma_y / (\Gamma_x \cup \{x,y\})| = K_n-m-1
\label{eq:Gamma_cardinalities}
\end{equation}
This situation is depicted in Figure \ref{fig:p2}.

Now, using arguments similar to those that lead to (\ref{eq:secondp1final}), we get 
\begin{eqnarray}
\lefteqn{P_{21}(n;\theta_n;m,\ell)} &&
\nonumber \\
 &=& \sum\limits_{i,j = 0}^{\ell-1}{K_n \choose i}{{K_n - 1} \choose j}  p_n^{i+j}(1-p_n)^{2K_n-i-j-1} ~\sum\limits_{i_1,j_1 = 0}^{\ell-1-i,\ell-1-j}{K_n-m -1\choose i_1}{K_n-m \choose j_1} 
\nonumber\\
& &~  \cdot  \left(\frac{p_nK_n}{n-1} \right)^{i_1+j_1}\left(1-\frac{p_nK_n}{n-1} \right)^{2K_n-2m-i_1-j_1-1}~ \sum\limits_{u = 0}^{\ell-1-\max{(i+i_1,j+j_1)}}{n-2K_n+m-1 \choose u}
\nonumber\\
& &~ \cdot (\bP{(z \sim x) \cap (z \sim y)})^u
{n-2K_n+m-1-u \choose \ell-1-i-i_1-u} (\bP{(z \sim x) \cap (z \sim y)^c}) ^{\ell-1-u-i-i_1} 
\nonumber\\
& &~ \cdot {n-2K_n+m-\ell+i+i_1 \choose \ell-1-j-j_1-u}
\cdot(\bP{(z \sim x)^c \cap (z \sim y)}) ^{\ell-1-u-j-j_1} 
\nonumber \\ 
& & ~ \cdot \left(\bP{(z \sim x)^c \cap (z \sim y)^c} \right)^{n-2K_n+m+1-2\ell+i+i_1+j+j_1+u}
\label{eq:secondp2final}
\end{eqnarray}
with $z$ denoting an arbitrary node in $\mathcal{V}/(\{x,y\}\cup \Gamma_x \cup \Gamma_y)$.
In (\ref{eq:secondp2final}), the following notation is used in the conditioning arguments: 
$d_x (\Gamma_x) = i$, $d_y (\Gamma_y / \{x\}) = j$, $d_x (\Gamma_y / (\Gamma_x \cup \{x,y\})) = i_1$, $d_y (\Gamma_x / (\Gamma_y \cup \{x,y\})) = j_1$, 
and $u$ denotes the number of nodes in $\mathcal{V}/(\{x,y\}\cup \Gamma_x \cup \Gamma_y)$ that are connected
to both $x$ and $y$; i.e., $u= |\{z \in \mathcal{V}/(\{x,y\}\cup \Gamma_x \cup \Gamma_y):~ (z\sim x)\cap(z \sim y)\}|$.

By direct comparison of (\ref{eq:secondp2final}) and (\ref{eq:secondp1final}), we find that
\begin{equation}
P_{21}(n;\theta_n;m,\ell) \leq (1+o(1)) (1-p_n)^{-1} \left(1-\frac{p_n K_n}{n-1}\right)^{-1} P_{1}(n;\theta_n;m,\ell - 1)
\label{eq:key_for_bounding_P21}
\end{equation}
as we note
\[
{K_n -1 \choose j} \leq {K_n \choose j} \qquad \textrm{and} \qquad {K_n -m-1 \choose i_1} \leq {K_n - m \choose i_1}
\]
and use the asymptotic equivalencies
\begin{eqnarray}
{n-2K_n+m-1 \choose u} &=& (1+o(1)) {n-2K_n+m-2 \choose u}
\nonumber \\
{n-2K_n+m-1-u \choose \ell -1-i-i_1-u} &=& (1+o(1)) {n-2K_n+m-2-u \choose (\ell -1)-i-i_1-u}
\nonumber \\
{n-2K_n+m-\ell + i+ i_1 \choose \ell -1-j-j_1-u} &=& (1+o(1)) {n-2K_n+m-2-(\ell-1)+i+i_1 \choose (\ell -1)-j-j_1-u}
\nonumber
\end{eqnarray}
that are immediate from the following arguments: Pick any positive constants $c_1$ and $c_2$, and recall that
$\lim_{n \to \infty} n-2K_n = \infty$ under the enforced assumptions. Then, for any $m=0,1,\ldots, K_n$, we have
\begin{eqnarray}
\frac{{n-2K_n+m \pm c_1 \choose c_2}}{{n-2K_n+m \choose c_2}} &=& \frac{(n-2K_n +m \pm c_1)!}{(n-2K_n+m \pm c_1 -c_2)!} \cdot \frac{(n-2K_n+m-c_2)!}{(n-2K_n+m)!}
\nonumber \\
&=& \frac{(n-2K_n+m\pm c_1) \cdots (n-2K_n+m\pm c_1 -c_2+1)}{(n-2K_n+m) \cdots (n-2K_n+m -c_2+1)}  
\nonumber \\
&=& \prod_{i=0}^{c_2-1} \left(1 \pm \frac{c_1}{n-2K_n+m -i} \right)
 \nonumber \\
&=& (1 \pm o(1))^{c_2}
\nonumber \\
&=& 1+o(1).
\label{eq:asymptotic_equiv_for_combinatorial}
\end{eqnarray}
 
We now report (\ref{eq:1_minus_pK_to_zero}) and
Proposition \ref{prop:second1} into (\ref{eq:key_for_bounding_P21}), and get
\begin{equation}
P_{21}(n;\theta_n;m,\ell) \leq (1+o(1)) (1-p_n)^{-1}\bP{D_{x,\ell-1}}^2, \qquad \ell=1,2,\ldots.
\label{eq:key_for_bounding_P21_B}
\end{equation}
Using the tight bound obtained for $\bP{D_{x,\ell}}$ in Proposition \ref{prop:for_key_1}, this then yields
\begin{equation}
P_{21}(n;\theta_n;m,\ell) \leq (1+o(1)) \bP{D_{x,\ell}}^2 (1-p_n)^{-1} \left(\frac{p_n K_n}{\ell} \left(1-\frac{K_n}{n-1} + \frac{1}{1-p_n} \right) \right)^{-2}
\label{eq:key_for_bounding_P21_C}
\end{equation}

The claim (\ref{eq:p_21_zero}) is now immediate as we note that  $(1-p_n)^{-1} = O(1)$ under the assumption that $\lim \sup_{n \to \infty} p_n < 1$ and 
\[
\lim_{n \to \infty} \left(\frac{p_n K_n}{\ell} \left(1-\frac{K_n}{n-1} + \frac{1}{1-p_n} \right) \right) = \infty
\]
since $p_n K_n = \Omega (\log n)$ from (\ref{eq:useful_p_n_K_n}) and we have $\left(1-\frac{K_n}{n-1} + \frac{1}{1-p_n} \right)  \geq 1$ from (\ref{eq:extra_term_lower_bound}).

We now consider the second term $P_{22}(n,\theta_n; m, \ell)$ in (\ref{eq:p2opened}). In view of (\ref{eq:p_21_zero}), it is clear that Proposition \ref{prop:second2} will be established if we show that
\begin{eqnarray}\label{eq:secondp22prop}
	P_{22}(n,\theta_n; m, \ell) \leq (1+ o(1))(1-p_n)^{-1}\bP{D_{x,\ell}}^2, \qquad \ell= 0,1,\ldots
\end{eqnarray}
for each $m=0,1,\ldots, K_n$ under the enforced assumptions.
We proceed as before and note that under the condition $(x \in \Gamma_y ~,~ y\notin \Gamma_x ~,~ |\Gamma_x \cup \Gamma_y| = 2K_n-m~,~ B_{xy}(p_n)=0)$, nodes $x$ and $y$ do {\em not} have an edge in between,  and both need to have $\ell$ neighbors among the $n-2$ nodes in $\mathcal{V}/\{x,y\}$. Otherwise, everything is just the same with the case 
of computing $P_{21}(n,\theta_n; m, \ell)$ including the set sizes given in (\ref{eq:Gamma_cardinalities}). Therefore, we have 
\begin{equation}
P_{22}(n,\theta_n; m, \ell) = P_{21}(n,\theta_n; m, \ell+1), \qquad \ell =0, 1, \ldots
\label{eq:relation_P_21_P_22}
\end{equation} 
for each $m=0,1,\ldots, K_n$. Now, we use (\ref{eq:key_for_bounding_P21}) in (\ref{eq:relation_P_21_P_22}) to get
\begin{eqnarray}
P_{22}(n,\theta_n; m, \ell)  &\leq& (1+o(1)) (1-p_n)^{-1} \left(1-\frac{p_n K_n}{n-1}\right)^{-1} P_{1}(n;\theta_n;m,\ell )
\nonumber \\
&\leq& (1+o(1)) (1-p_n)^{-1} \bP{D_{x,\ell}}^2,
\nonumber
\end{eqnarray}
where in the last step we used (\ref{eq:1_minus_pK_to_zero}) and
Proposition \ref{prop:second1}. This establishes (\ref{eq:secondp22prop}), and the proof of
Proposition \ref{prop:second2} is now complete in view of (\ref{eq:p_21_zero}) and (\ref{eq:p2opened}).
\myendpf

\section{A proof of Proposition \ref{prop:second3}}
\label{sec:second3}

We start as in the proof of Proposition \ref{prop:second2}
and condition on the event $(x \sim_{B} y)$ to get 
\begin{eqnarray}
	P_{3}(n,\theta_n; m, \ell) = p_nP_{31}(n,\theta_n; m, \ell) + (1-p_n)P_{32}(n,\theta_n; m, \ell)
\label{eq:p3opened}
\end{eqnarray}
where
\begin{eqnarray}
	P_{31}(n,\theta_n; m, \ell) &=& \bP{D_{x,\ell}\cap D_{y,\ell} ~\big |~ x\in \Gamma_y ~,~ y\in \Gamma_x ~,~ |\Gamma_x \cup \Gamma_y| = 2K_n-m ~,~ B_{xy}(p_n)=1}
\label{eq:p31} \nonumber
	\\
		P_{32}(n,\theta_n; m, \ell) &=& \bP{D_{x,\ell}\cap D_{y,\ell} ~\big|~ x\in \Gamma_y ~,~ y\in \Gamma_x ~,~ |\Gamma_x \cup \Gamma_y| = 2K_n-m ~,~ B_{xy}(p_n)=0}
\label{eq:p32} \nonumber
\end{eqnarray}

\begin{figure}[!t]
\centering
\includegraphics[totalheight=0.22\textheight,
width=.5\textwidth]{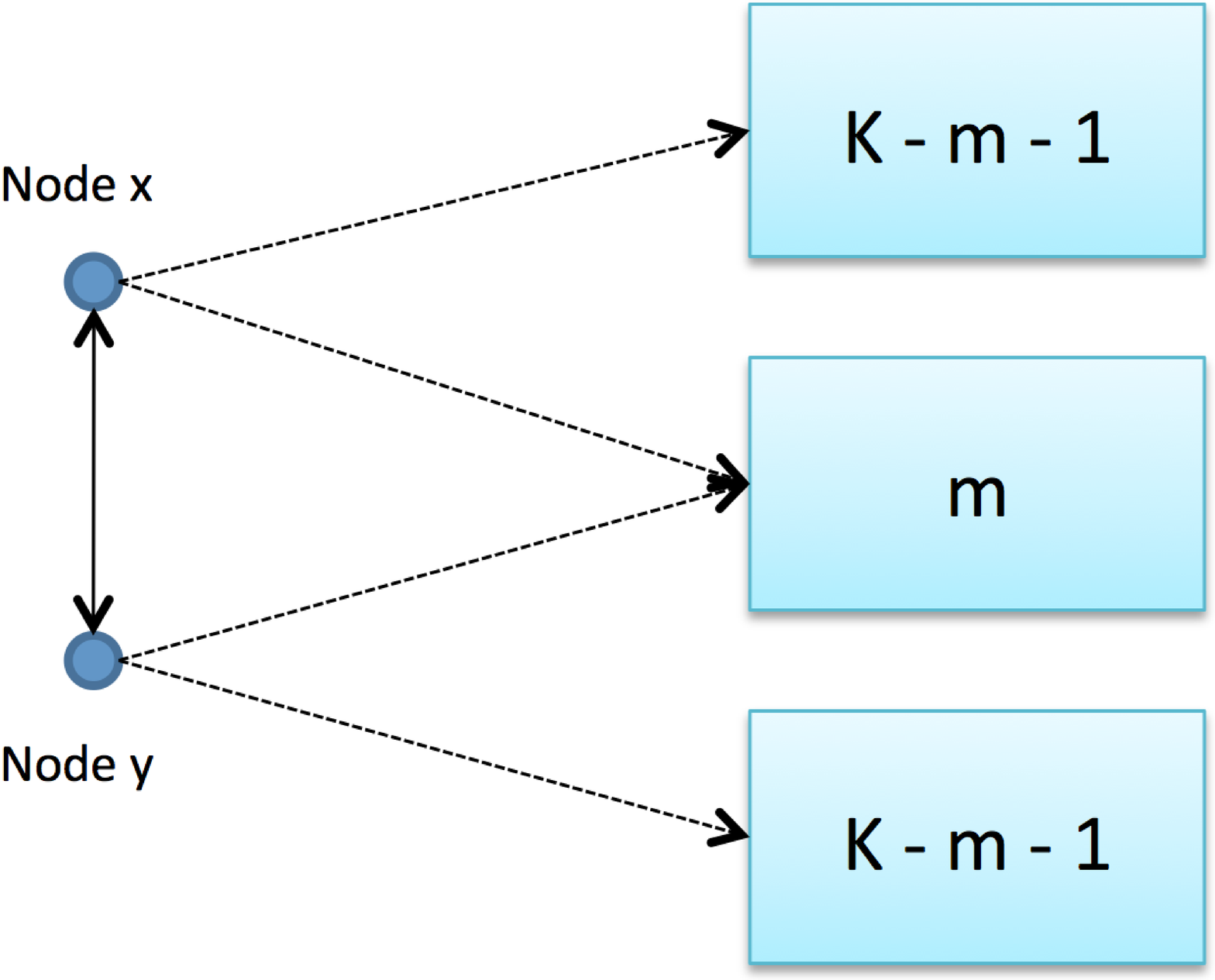} 
\caption{\sl Depicting the condition 
$(x \in \Gamma_y ~,~ y\in \Gamma_x ~,~ |\Gamma_x \cup \Gamma_y| = 2K_n-m ~,~ B_{xy}=1)$ for calculating
$P_{31}(n,\theta_n; m, \ell)$. Dashed lines emanating from
a node $x$ stand for the set of nodes in $\Gamma_x$ and a dashed line between $x$ and $y$
is made bold if $B_{xy}=1$.}
\label{fig:p3}
\end{figure}

In what follows, we will compute $P_{31}(n,\theta_n; m, \ell)$ and $P_{32}(n,\theta_n; m, \ell)$ in turn. 
We will start by showing that
\begin{equation}
P_{31}(n,\theta_n; m, \ell) = o\left(\bP{D_{x,\ell}}^2 \right), \qquad \ell = 1, 2, \ldots 
\label{eq:p_31_zero}
\end{equation}
for each $m=0,1,\ldots, K_n$ under the enforced assumptions. To do so, we note that the condition 
$(x \in \Gamma_y ~,~ y\in \Gamma_x ~,~ |\Gamma_x \cup \Gamma_y| = 2K_n-m~,~ B_{xy}(p_n)=1)$ amounts to having \begin{equation}
|\Gamma_x \cap \Gamma_y| = m, \quad |\Gamma_x / (\Gamma_y \cup \{x,y\})| = K_n-m -1, \quad \textrm{and} \quad |\Gamma_y / (\Gamma_x \cup \{x,y\})| = K_n-m-1
\label{eq:Gamma_cardinalities_2}
\end{equation}
Also under this condition, nodes 
$x$ and $y$ do have an edge in between, and they just need to have $\ell-1$ additional neighbors among the $n-2$ nodes in $\mathcal{V}/\{x,y\}$. These facts are depicted in Figure \ref{fig:p3}.

Now, using arguments similar to those that lead to (\ref{eq:secondp1final}) and (\ref{eq:secondp2final}), 
we get 
\begin{eqnarray}
\lefteqn{P_{31}(n;\theta_n;m,\ell)} &&
\nonumber \\
 &=& \sum\limits_{i,j = 0}^{\ell-1}{K_n -1 \choose i}{{K_n - 1} \choose j}  p_n^{i+j}(1-p_n)^{2K_n-i-j-2} ~\sum\limits_{i_1,j_1 = 0}^{\ell-1-i,\ell-1-j}{K_n-m -1\choose i_1}{K_n-m-1 \choose j_1} 
\nonumber\\
& &~  \cdot  \left(\frac{p_nK_n}{n-1} \right)^{i_1+j_1}\left(1-\frac{p_nK_n}{n-1} \right)^{2K_n-2m-i_1-j_1-2}~ \sum\limits_{u = 0}^{\ell-1-\max{(i+i_1,j+j_1)}}{n-2K_n+m \choose u}
\nonumber\\
& &~ \cdot (\bP{(z \sim x) \cap (z \sim y)})^u
{n-2K_n+m-u \choose \ell-1-i-i_1-u} (\bP{(z \sim x) \cap (z \sim y)^c}) ^{\ell-1-u-i-i_1} 
\nonumber\\
& &~ \cdot {n-2K_n+m-\ell+1+i+i_1 \choose \ell-1-j-j_1-u}
\cdot(\bP{(z \sim x)^c \cap (z \sim y)}) ^{\ell-1-u-j-j_1} 
\nonumber \\ 
& & ~ \cdot \left(\bP{(z \sim x)^c \cap (z \sim y)^c} \right)^{n-2K_n+m+2-2\ell+i+i_1+j+j_1+u}
\label{eq:secondp3final}
\end{eqnarray}
with $z$ denoting an arbitrary node in $\mathcal{V}/(\{x,y\}\cup \Gamma_x \cup \Gamma_y)$.
The notation used in the conditioning arguments of (\ref{eq:secondp3final}) are as follows:
$d_x (\Gamma_x) = i$, $d_y (\Gamma_y / \{x\}) = j$, $d_x (\Gamma_y / (\Gamma_x \cup \{x,y\})) = i_1$, $d_y (\Gamma_x / (\Gamma_y \cup \{x,y\})) = j_1$, 
and $u$ denotes the number of nodes in $\mathcal{V}/(\{x,y\}\cup \Gamma_x \cup \Gamma_y)$ that are connected
to both $x$ and $y$; i.e., $u= |\{z \in \mathcal{V}/(\{x,y\}\cup \Gamma_x \cup \Gamma_y):~ (z\sim x)\cap(z \sim y)\}|$.

By direct comparison of (\ref{eq:secondp3final}) and (\ref{eq:secondp2final}), we find that
\begin{equation}
P_{31}(n;\theta_n;m,\ell) \leq (1+o(1)) (1-p_n)^{-1} \left(1-\frac{p_n K_n}{n-1}\right)^{-1} P_{21}(n;\theta_n;m,\ell)
\label{eq:key_for_bounding_P31}
\end{equation}
upon noting that 
\[
{K_n -1 \choose i} \leq {K_n \choose i} \qquad \textrm{and} \qquad {K_n -m-1 \choose j_1} \leq {K_n - m \choose j_1}
\]
and using the bounds
\begin{eqnarray}
{n-2K_n+m \choose u} &=& (1+o(1)) {n-2K_n+m-1 \choose u}
\nonumber \\
{n-2K_n+m-u \choose \ell -1-i-i_1-u} &=& (1+o(1)) {n-2K_n+m-1-u \choose \ell -1-i-i_1-u}
\nonumber \\
{n-2K_n+m-\ell +1+ i+ i_1 \choose \ell -1-j-j_1-u} &=& (1+o(1)) {n-2K_n+m-\ell+i+i_1 \choose \ell -1-j-j_1-u}
\nonumber
\end{eqnarray}
that are immediate from (\ref{eq:asymptotic_equiv_for_combinatorial}). The claimed result (\ref{eq:p_31_zero}) follows immediately by reporting (\ref{eq:p_21_zero}) into (\ref{eq:key_for_bounding_P31}) and noting that 
\begin{equation}
(1-p_n)^{-1} \left(1-\frac{p_n K_n}{n-1}\right)^{-1} = O(1)
\end{equation}
under the enforced assumptions; just recall that $\lim \sup_{n \to \infty} p_n < 1$ and use (\ref{eq:1_minus_pK_to_zero}).

In view of (\ref{eq:p_31_zero}) and (\ref{eq:p3opened}), Proposition \ref{prop:second3} will be established if we show that 
\begin{eqnarray}\label{eq:secondp32prop}
	P_{32}(n,\theta_n; m, \ell) \leq (1+ o(1))(1-p_n)^{-2}\bP{D_{x,\ell}}^2, \qquad \ell=0,1,\ldots
\end{eqnarray}
for each $m=0,1,\ldots, K_n$.
In calculation of $P_{32}(n,\theta_n; m, \ell)$, we need to consider the condition $(x \in \Gamma_y ~,~ y\in \Gamma_x ~,~ |\Gamma_x \cup \Gamma_y| = 2K_n-m~,~ B_{xy}(p_n)=0)$, where nodes $x$ and $y$ do {\em not} have an edge in between, and both need to have $\ell$ neighbors among the $n-2$ nodes in $\mathcal{V}/\{x,y\}$. This is the only difference between the probabilities $P_{31}(n,\theta_n; m, \ell)$ and $P_{32}(n,\theta_n; m, \ell)$. Except this difference, 
all statistical equivalencies and relations are the same 
including the set sizes given in (\ref{eq:Gamma_cardinalities_2}). Therefore, it is immediate that
\begin{equation}
P_{32}(n,\theta_n; m, \ell) = P_{31}(n,\theta_n; m, \ell+1), \qquad \ell =0, 1, \ldots
\label{eq:relation_P_31_P_32}
\end{equation} 
for each $m=0,1,\ldots, K_n$. We now use (\ref{eq:key_for_bounding_P31}) in (\ref{eq:relation_P_31_P_32}) to get
\begin{eqnarray}
P_{32}(n,\theta_n; m, \ell)  &\leq& (1+o(1)) (1-p_n)^{-1} \left(1-\frac{p_n K_n}{n-1}\right)^{-1} P_{21}(n;\theta_n;m,\ell + 1)
\nonumber \\
&\leq& (1+o(1)) (1-p_n)^{-2} \cdot \bP{D_{x,\ell}}^2,
\nonumber
\end{eqnarray}
where in the last step we used the previously obtained bound (\ref{eq:key_for_bounding_P21_B}) together with (\ref{eq:1_minus_pK_to_zero}). This establishes (\ref{eq:secondp32prop}), and the proof of
Proposition \ref{prop:second3} is now complete in view of (\ref{eq:p_31_zero}) and (\ref{eq:p3opened}).
\myendpf

\bibliographystyle{abbrv}
\bibliography{related}

\end{document}